\documentclass[reprint,aps,showpacs,amsmath,amssymb,superscriptaddress]{revtex4-1}%

\usepackage{graphicx}
\usepackage{dcolumn}
\usepackage{bm}
\usepackage{hyperref}
\usepackage{color}
\usepackage{natbib}

\newcommand{\rmd}{{\rm d}}
\newcommand{\rme}{{\rm e}}
\newcommand{\rmi}{{\rm i}}

\newcommand{\Ud}[1]{\hspace{-0.5ex}\mathrm{d}{#1}\;}

\begin{document}

\title{Nonequilibrium Mean-Field  Theory of Resistive Phase Transitions}
\author{Jong E. Han}
\email{jonghan@buffalo.edu}
\affiliation{Department of Physics, State University of New York at Buffalo, Buffalo, New York 14260, USA}
\author{Jiajun Li}
\affiliation{Department of Physics, State University of New York at Buffalo, Buffalo, New York 14260, USA}
\affiliation{Department of Physics, University of Erlangen-Nuremberg, Erlangen, Germany}
\author{Camille Aron}
\affiliation{Laboratoire de Physique Th\'eorique, \'Ecole Normale Sup\'erieure, CNRS, PSL University, Sorbonne Universit\'e, Paris 75005, France}
\affiliation{Instituut voor Theoretische Fysica, KU Leuven, Belgium}
\author{Gabriel Kotliar}
\affiliation{Department of Physics, Rutgers University, New Jersey 08854, USA}

\date{\today}

\begin{abstract} 

We investigate the quantum mechanical origin of resistive phase
transitions in solids driven by a constant electric field in the
vicinity of a metal-insulator transition.  We perform a nonequilibrium
mean-field analysis of a driven-dissipative anti-ferromagnet, which we
solve analytically for the most part.  We find that the
insulator-to-metal transition (IMT) and the metal-to-insulator
transition (MIT) proceed by two distinct electronic mechanisms:
Landau-Zener processes, and the destabilization of
metallic state by Joule heating, respectively.
However, we show that
both regimes can be unified in a common effective thermal description,
where the effective temperature $T_{\rm eff}$ depends on the state of
the system.  This explains recent experimental measurements in which the
hot-electron temperature at the IMT was found to match the equilibrium
transition temperature.  Our analytic approach enables us to formulate
testable predictions on the non-analytic behavior of $I$-$V$ relation
near the insulator-to-metal transition. Building on these successes, we
propose an effective Ginzburg-Landau theory which paves the way to
incorporating spatial fluctuations, and to bringing the theory closer to
a realistic description of the resistive switchings in correlated
materials.

\end{abstract}


\maketitle

\section{Introduction}

Phase transitions driven by out-of-equilibrium conditions is one of the
most fascinating and challenging topics of modern condensed matter.  The
phenomenon of resistive switching (RS) refers to the sudden massive drop
of resistivity experienced by many insulating materials when subject to
a voltage bias or to an electric field. Importantly, the
metal-to-insulator transition (MIT) on an up-sweep of the electric field
takes place at a lower threshold field than the insulator-to-metal
transition (IMT) on the down-sweep, resulting in hysteretic $I-V$
characteristics.  The growing interest for this phenomenon over the last
decades has been stimulated by the perspective of designing logic
devices for digital computation~\cite{janod,stoliar,jslee,ridley}.  In
particular, memristor physics has turned into a full-blown research
effort to create novel reliable non-volatile logic devices. Very
recently,  RS in Mott insulators has been proposed to realize the
neurons that make up artificial neural networks: two-terminal devices
that can reproduce the fast spiking of neurons when subject to electric
pulses~\cite{Cario}.

In addition to
its appeal for applied physics, resistive switching is a fundamental
physics problem, as a prototypical nonequilibrium phase transition of
quantum many-body systems.  Despite its importance, the theoretical
understanding of RS  has remained unsatisfactory. A rather successful
heuristic approach is the resistor network
theory~\cite{dubson,driscoll,guiot,janod}, which models the materials by
a classical network of resistors with empirical electric and thermal
properties, and where an electric filament can percolate across the
insulating matrix.  However, the vast diversity of the systems
displaying RS, from intrinsic semiconductors to transition metal
compounds~\cite{janod}, possibly through various microscopic mechanisms,
together with the formidable theoretical difficulty in solving the
nonequilibrium dynamics of quantum many-body  systems, are to be blamed
for our current lack of a unifying quantum theory of RS. It is only
recently that the community has started developing the methodologies to
combine strong electronic interactions and  nonequilibrium
drives~\cite{aoki}.

In the past few decades, the theory of quantum nonequilibrium dynamics
in general has made important progress. Far-from-equilibrium transport
theory has found countless applications in nano-junctions, based on the
Landauer-B\"uttikker formalism~\cite{datta_book}. Recently, stimulated
by progress in ultrafast measurement techniques~\cite{jager}, the
relaxation dynamics of electrons at the femto-second scale has been
extensively studied in solids and optical lattices~\cite{kemper}. The
general idea behind our work is rather to understand how the electronic
state continuously evolves away from equilibrium when a  steady finite
electric field is adiabatically turned on.  The theoretical studies of
quantum phase transitions of nonequilibrium steady states is still
limited in solids.  Perturbative studies starting from a metallic state
under a DC field~\cite{ybkim,mitra,li_prl} have exposed the importance
of Joule-heating, whereby the electric field acts as an effective
temperature. This has lead to classify RS in the same universality class
as the continuous Ising transition that characterizes the equilibrium
paramagnetic-to-anti-ferromagnetic transition at the N\'eel temperature.
In contrast to the previous efforts, we investigate the
insulator-to-metal RS and find a discontinuous nonequilibrium phase
transition, in stark differences with the Ising class.

We study here insulating transition-metal oxides or 
transition-metal chalcogenides with a relatively small bandgap,
$\Delta_0  \lesssim 1$~eV, and for which the measured switching fields
are in the range of $E_{\rm IMT} \sim 1-10$~kV/cm.  RS in those
correlated insulators poses two major puzzles: (1) the typical switching
field (or voltage drop per unit cell) is sub-meV, much smaller than the
bandgap, therefore incapable of turning the insulator band structure
into a metal, (2) there is a controversy over the nature of the
underlying mechansim: electronic~\cite{mayer,jager,mazza} vs
thermal~\cite{zimmers,guenon,duchene,sujay} scenarios.  The electronic
scenarios support the idea that the RS is due to the electric-field
driven acceleration of the electrons which triggers a sudden change of
the electronic transport properties. Various ideas such as the formation
of in-gap states~\cite{aron2012,lee2014}, Landau-Zener
tunneling~\cite{oka2003,oka2005}, avalanches of impact ionization
events~\cite{guiot}, and multi-band interacting model~\cite{mazza} have
been proposed to resolve the aforementioned energy-scale problem.  On
the other hand, the thermal scenarios support the idea that the
electronic current created by the electric field causes an overhaul
temperature increase via Joule-heating, essentially bringing the system
to undergo a thermally-driven equilibrium phase transition rather than a
truly nonequilibrium phase transition.  Such a mechanism would be
effective in overcoming the large energy gap discussed above, but it is
considered to require a long time to build up the necessary temperature,
in contradiction with the fast switching times of RS that are observed
experimentally.  Altogether, the experimental evidences give partial
support to each scenario and the debates between the two camps have
remained inconclusive for decades.

In this work, we analytically elucidate the above puzzles and explain
how the electronic and thermal scenarios are in fact different sides of
the same coin, by solving explicitly the case of an ordered insulator
driven by an electric field. The scenario we uncover consists in the
electric field $E$ effectively coupling to the order parameter $\Delta$
via a state-dependent effective temperature, $T_{\rm eff}(\Delta)$.
Ultimately, this sets the small energy scale of the switching fields,
and yields testable predictions on the critical scaling of the $I-V$
curves at the IMT.

We work with a model of a driven-dissipative quantum anti-ferromagnet
that we have recently identified in Ref.~\cite{nanolett} as a minimal
model for RS. A similar model had already been introduced and studied in
the pioneering work of Sugimoto \textit{et al.}~\cite{sugimoto}. The
numerical study of the nonequilibrium steady states in
Ref.~\cite{nanolett} showed that it reproduced most of the feature of RS
which are observed experimentally, such as the existence of a
bi-stability region between of the metallic and insulating solutions,
the S-shaped $I$-$V$ characteristics, the formation of hot metallic
filaments across the sample whose dynamics are responsible for a
negative differential-resistance~\cite{ridley,htkim,Kim:2010kc}.
Although much insight could be gained from the numerics, a comprehensive
and unambiguous analytic understanding for the inner workings of the
results has been lacking.

The paper is organized as follows.  In Section II, we start off with a
simple single-band metal subject to an electric field and dissipating
heat in a zero-temperature bath. We compute the Keldysh Green's
functions (GFs) in the nonequilibrium steady state, and we obtain an
explicit expression of the nonequilibrium distribution function. We then
generalize the approach to a driven-dissipative anti-ferromagnet. The
corresponding GFs are derived by means of a mean-field approximation
where the order parameter is taken to be the charge gap, $\Delta$.  In
Section III, we analyze the insulating solutions in both the small and
large gap regimes, characterizing the nonequilibrium excitations by
means of an effective temperature $T_{\rm eff}(\Delta)$.  In Section IV,
after a brief review of the equilibrium case, we derive and solve the
self-consistent equation on the nonequilibrium mean-field order
parameter, and we identify the switching fields of the
insulator-to-metal and the metal-to-insulator transitions, $E_{\rm IMT}$
and $E_{\rm MIT}$, respectively. Since most RS experiments are realized at 200-300~K, 
we later generalize our results to
finite-temperature baths. In Section V, we reformulate the
nonequilibrium mean-field theory in terms of an effective free energy
$\mathcal{F}(\Delta)$.  In Section IV, we conclude and give additional
discussions.

\section{Quantum Nonequilibrium Formulation}\label{sec:formulation}
We first present our analytical approach, how we incorporate the nonequilibrium drive and the dissipation, within the case of a non-interacting single-band metal. We later move to the more complex case of a correlated anti-ferromagnet. While we limit our discussions to one-dimensional models, most of our conclusions are also valid in higher dimensions as long as low-dimensional correlation effects remain unimportant.

\subsection{Elementary Case: Single-Band Metal} \label{sec:metal}
Let us consider a tight-binding model of electrons set in motion by a DC electric field $E$. To prevent the Joule effect from heating up the sample to very high temperatures, and from eventually completely melting the system, it is necessary to couple it  to a large environment that can effectively dissipate its excess energy. 
As a rudimentary mechanism, we employ simple thermal bath of fermions which create Ohmic dissipation and satisfy the basic requirements consistent with the Boltzmann transport theory~\cite{prb2013a,prb2013b,li_prl}.
Besides  dissipation, the baths are also a crucial element because they allow to explore the RS in finite temperature environments, and thus to make the connection with experiments.
We first set up the problem on a one-dimensional lattice, then later linearize the dispersion relation to work with a continuum version.

\subsubsection{Lattice Model}

The total Hamiltonian of a simple metallic chain reads~\cite{prb2013a}
$\hat{H}_{\rm tot} =  \hat{H} + H_{\rm bath}$ with 
\begin{align}
\hat{H} & \!=\!  -t\sum_\ell(d^\dagger_{\ell+1}d_\ell+{\rm H.c.})  -E\sum_\ell\ell d^\dagger_\ell d_\ell, \\
\hat{H}_{\rm bath} & \!= \!  \sum_{\ell\alpha} \! ( \epsilon_\alpha - E \ell ) c^\dagger_{\ell\alpha}c_{\ell\alpha} \!
 -\! \frac{g}{\sqrt{L}} \! \sum_{\ell\alpha}(c^\dagger_{\ell\alpha} d_\ell+ {\rm H.c.}),
\end{align} 
where $d^\dagger_{\ell}$ is the creation operator of an electron at site
$\ell$, and $c^\dagger_{\ell\alpha}$ the creation of an electron in the
fermion bath coupled to site $\ell$, with the continuum index $\alpha$.
We set the lattice constant $a=1$, the electric charge $e=1$, and
$\hbar=1$.  The coupling between the orbital at site $\ell$ and its
local bath is given by the coupling constant $g$, and it yields the
local hybridization function $\Gamma(\omega)=(g^2/L)\sum_\alpha
(\omega-\epsilon_\alpha)$. We assume the baths to be identical at all
sites, and with a structureless spectrum such that
$\Gamma(\omega)=\Gamma$.

The DC electric field is incorporated in the Coulomb gauge via the static electric potential $-E\ell$.
For simplicity, we consider $E \geq 0$. In this gauge, the thermal
statistics of the bath degrees of freedom, the $c_{\ell \alpha}$'s, is given by the Fermi-Dirac distribution function where the original zero chemical potential is shifted by $-E\ell$ at site~$\ell$,
\begin{equation}
f_0(\omega+\ell E)=[\rme^{-(\omega+\ell E)/T_{\rm b}}+1]^{-1},
\end{equation}
and where $T_{\rm b}$ is the bath temperature. In the following, we consider a zero-temperature environment by setting $T_{\rm b}=0$, except in Section~\ref{sec:finiteT}.
Within the Keldysh Green's function formalism, the dissipation by the fermion
baths is exactly incorporated in the retarded and lesser self-energies
at site $\ell$ as
\begin{equation}
\Sigma^r_\ell(\omega) = -\rmi\Gamma,\quad
\Sigma^<_\ell(\omega) = 2\rmi\Gamma f_0(\omega+\ell E),
\label{hybr}
\end{equation}
respectively.
One defines the retarded and lesser Green's functions, $G^{r}_{ij}(\omega)$ and  $G^{<}_{ij}(\omega)$, respectively, as
\begin{eqnarray}
G^r_{ij}(t,t') & = & -\rmi\Theta(t-t')\langle
\{d_i(t),d^\dagger_j(t')\}\rangle,\\
G^<_{ij}(t,t') & = & \rmi\langle d^\dagger_j(t')d_i(t)\rangle.
\end{eqnarray}
Once the steady state has been reached, the Green's functions are time-translational invariant (though they are not space translational invariant due to our choice of gauge).
Using Dyson's equation on the lesser Green's function, its local component can then be computed as 
\begin{eqnarray}
G^<_{\rm loc}(\omega) & = &
\sum_{\ell=-\infty}^{+\infty}G^r_{0\ell}(\omega)\Sigma^<_\ell(\omega)
G^r_{0\ell}(\omega)^* \nonumber
\\
& = & 
2\rmi\Gamma\sum_{\ell=-\infty}^{+\infty}|G^r_{0\ell}(\omega)|^2f_0(\omega+\ell E).
\label{gless}
\end{eqnarray}
This problem has been solved numerically in Ref.~\cite{prb2013b}, which led to identification of an effective temperature for the electrons driven by a small  electric field and coupled to a zero-temperature bath
\begin{equation}
T_{\rm eff}=\frac{\sqrt{6}}{\pi}\frac{tE}{\Gamma}.
\label{teff0}
\end{equation}
While a current carrying steady state cannot be strictly considered as a thermal state, this simple characterization of the electronic excitations by a finite temperature proportional to $E/\Gamma$, \textit{i.e.} drive over dissipation,  nevertheless exposes clearly the driven-dissipative nature of the electronic steady state.

\subsubsection{Continuum Model}
In this paper, we work in the continuum limit where analytic approaches become more amenable. To this end, we linearize the
tight-binding dispersion relation near the chemical potential. Setting aside the dissipation for a moment, we obtain the Hamiltonian
\begin{equation}
\hat{H} = \sum_{\lambda=\pm} 
\int \Ud{x}
\psi^\dagger_\lambda(x) {h}_{\lambda}(x) \psi_\lambda(x),
\end{equation}
where  $\psi_\lambda(x)$ is the electron field operator of right ($\lambda = +$) and left ($\lambda = -$) movers evolving according to the Hamilonian density
\begin{align} \label{eq:ho1}
{h}_{\lambda}(x) =
-\rmi\lambda
v_0\partial_x -Ex,
\end{align}
and $v_0 > 0$ is the group velocity. 
In our numerics, we use $(\hbar/a)v_0$ as the unit of energy by setting it to unity.
Re-incorporating the dissipation by using the hybridization to the baths in Eq.~(\ref{hybr}), the Dyson equation
for the retarded GF reads
\begin{equation}
(\rmi \partial_t - h_{\lambda}(x)  +\rmi\Gamma)G^r_\lambda(x,x';t)=\delta(t)\delta(x-x'),
\end{equation}
whose solution can be expressed in the spectral representation as
\begin{equation}
G^r_\lambda(x,x';\omega)=\int\frac{\phi_\lambda(x,\omega')\phi^*_\lambda(x',\omega')}{
\omega-\omega'+\rmi\Gamma}\frac{\rmd\omega'}{2\pi v_0},
\label{gret}
\end{equation}
where $\phi_\lambda(x,\omega)$ is the eigen-function of the
dissipationless Hamiltonian in Eq.~(\ref{eq:ho1}) at energy $\omega$, \textit{i.e.}
\begin{equation} \label{eq:phi0}
h_{\lambda}(x) \phi_\lambda(x,\omega)=\omega \, \phi_\lambda(x,\omega).
\end{equation}
The continuum version of the local lesser GF given in Eq.~(\ref{gless}), with a bath temperature $T_{\rm b} = 0$, now reads
\begin{equation}
G^<_{\rm loc}(\omega)
=2\rmi\Gamma \int_{-\infty}^{-\omega/E}\frac12\sum_\lambda|G^r_\lambda(0,x;\omega)|^2 \rmd x.
\label{lesser0}
\end{equation}
The local energy distribution function $f(\omega)$ can be accessed via
\begin{equation} 
f(\omega)=-\frac{G^<_{\rm loc}(\omega)}{2 \rmi \,\mbox{Im }G^r_{\rm loc}(\omega)}.
\label{feff0}
\end{equation}
In equilibrium (at $E=0$), the fluctuation-dissipation theorem between retarded and lesser GFs ensures that the energy distribution is governed by the usual zero-temperature Fermi-Dirac distribution.
Out of equilibrium ($E > 0$), one simple way to quantify the amount of nonequilibrium excitations around the chemical potential is to introduce an effective temperature, $T_{\rm eff}$. In regimes with relatively few excitations concentrated around the chemical potential, it is quite convenient to use the following definition of the effective temperature based on the Sommerfeld expansion~\cite{nanolett},
\begin{equation}
T_{\rm eff}^2=\frac{6}{\pi^2}\int_{-\infty}^\infty\omega[f(\omega)
-\Theta(-\omega)]\rmd\omega,
\label{teff}
\end{equation}
and which is consistent with the equilibrium temperature when $f(\omega)$ is the
Fermi-Dirac distribution.

\subsubsection{Analytic Solution}

Owing to the linearized dispersion relation, $G^r_\lambda(x,x';\omega)$ and $G^<_{\rm loc}(\omega)$ can be computed explicitly. Indeed, 
the Schr\"odinger equation in Eq.~(\ref{eq:phi0}) has a simple solution reading
\begin{equation} 
\phi_\lambda(x,\omega) = \exp\left[\frac{\rmi\lambda}{v_0}\left(\omega x+\frac12
Ex^2\right)\right].
\label{phi0}
\end{equation}
After performing a contour integral in Eq.~(\ref{gret}), we obtain
\begin{equation}
G^r_\lambda(x,x';\omega)=-\frac{\rmi}{v_0}\Theta(\lambda(x-x'))\rme^{
-\frac{\Gamma}{v_0}|x-x'|}\rme^{\rmi\lambda\varphi},
\label{g+0}
\end{equation}
with the phase $\varphi=[\omega(x-x')+\frac12 E(x^2-x'^2)]/v_0$.
The local retarded GF reads 
\begin{align} \label{eq:grloc}
G^r_{\rm loc}(\omega)=-\rmi/(2v_0).
\end{align}
Note that, unlike in the lattice calculations, the
spectral function $-\pi^{-1}\mbox{Im }G^r_{\rm loc}(\omega)=1/(2\pi v_0)$
does not feature Bloch-Zener peaks equally spaced in energy by $e E a$,
due to the lack of a finite lattice constant in the continuum model.

Using Eqs.~(\ref{lesser0}), (\ref{feff0}) and~(\ref{eq:grloc}), the local lesser GF reads
\begin{equation} \label{eq:glessloc}
G^<_{\rm loc}(\omega)=\frac{\rmi}{v_0} f(\omega),
\end{equation}
with the local energy distribution function
\begin{equation}
f(\omega)=\left\{
\begin{array}{ll}
\frac12\rme^{-2\Gamma \omega/(v_0 E)} & \mbox{for }\omega>0 \\
1-\frac12\rme^{2\Gamma \omega/(v_0 E)} & \mbox{for }\omega<0
\end{array}
\right..
\label{distr0}
\end{equation}
This expression is in agreement with the quantum Boltzmann theory of
Mitra and Millis~\cite{mitra}.

The above expression for $f(\omega)$ shows that the steady-state carries nonequilibrium excitations above the chemical potential, on an energy scale controlled by $v_0 E / \Gamma$.  More quantitatively, 
using Eq.~(\ref{teff}), it corresponds to an effective temperature
\begin{align} \label{eq:teff0}
T_{\rm eff} = \sqrt{\frac{3}{2}} \frac{v_0 E}{\pi \Gamma},
\end{align}
which agrees with the expression in Eq.~(\ref{teff0}) that was obtained using linear response theory in the half-filled lattice model with $v_0=2t$~\cite{prb2013b}.

\subsection{Driven-Dissipative Anti-Ferromagnet}
We now turn to the case of the anti-ferromagnet.
We consider a staggered phase, where the one-dimensional lattice is split in two sublattice, $A$ and $B$, the  energy levels of which are alternating by $\pm \Delta$ with $\Delta \geq 0$.
While in this Section, the value of $\Delta$ is considered arbitrary, it can be seen as originating from a mean-field treatment of a local interaction between the electrons. This will be the topic of the next Section, where the value of $\Delta$ will be set self-consistently and the emergence of anti-ferromagnetism will be studied systematically via a mean-field approach.

\subsubsection{Continuum Model}
Setting aside the dissipation for a moment, we consider the continuous Hamiltonian 
\begin{align}
\hat H = \int \Ud{x}   \phi^\dagger(x) h(x) \phi(x),
\end{align}
with the local fermion degrees of freedom $\phi(x) \equiv (\phi_A(x),\phi_B(x))^{T}$ and the two-band Hamiltonian density 
\begin{equation}
h(x) =
\left(\begin{array}{cc}
-\Delta-Ex & -\rmi v_0\partial_x \\
-\rmi v_0\partial_x & \Delta-Ex
\end{array}\right).
\label{h0}
\end{equation}
It is useful to work with the rotated wavefunctions $\phi_\pm=1/\sqrt{2}(\phi_A\pm\phi_B)$ by performing  a unitary transformation 
\begin{equation}
\hat{U}=\frac{1}{\sqrt{2}}\left(\begin{array}{cc} 1 & 1 \\ 1 & -1\end{array}\right).
\end{equation}
In this basis, the dissipationless Schr\"odinger equation reads
\begin{equation}
\left(\begin{array}{cc}
\omega+\rmi v_0\partial_x + Ex & \Delta \\
\Delta & \omega-\rmi v_0\partial_x +Ex
\end{array}\right)\left(\begin{array}{c}
\phi_+ \\ \phi_-
\end{array}\right)=0.
\label{diff1}
\end{equation}
When $\Delta=0$, $\phi_\pm$ satisfy the same differential equations as
in the previous single electronic band of left- and right-movers.
Therefore, we parametrize the solutions
$\phi^\lambda=(\phi^\lambda_+,\phi^\lambda_-)^T$ of the above
equations by the superscript $\lambda = L, R$. This problem can be
understood as the Schwinger effect~\cite{schwinger51} where
particle-hole pairs are created from the one-dimensional massive Dirac
field (with mass $\Delta$) by a static electric field.

Once these eigen-functions of the dissipationless Hamiltonian are computed (see below), they can be used to construct the 
GFs in the presence of dissipation. The retarded GF is given by ($a,b=\pm$)
\begin{equation}
G^r_{ab}(x,x';\omega)=\int\frac{\rho_{ab}(x,x';\omega')}{
\omega-\omega'+\rmi\Gamma}\frac{\rmd\omega'}{2\pi v_0},
\label{retardg}
\end{equation}
with the dissipationless spectral function
\begin{equation}
\rho_{ab}(x,x';\omega)=\sum_{\lambda=R,L}
\phi^\lambda_a(x,\omega)\phi^\lambda_b(x',\omega)^*.
\label{spectral}
\end{equation}
Generalizing Eq.~(\ref{lesser0}) to a two-band electronic structure, the lesser GF at $x=x'$ is given by
\begin{equation}
G^<_{ab}(\omega)=2\rmi\Gamma\sum_{c=\pm}\int_{-\infty}^{-\omega/E} \!\!\!\!\!\! \rmd x \,G^r_{ac}(0,x;\omega)
G^r_{bc}(0,x;\omega)^*,
\label{lesserg}
\end{equation}
where we recall that the bath temperature is set to zero.
We define the local retarded and lesser GFs as equal-weight averages of the A and B sublattice,
\begin{equation}
G_{\rm loc}^{r/<}(\omega) = \frac{G_{AA}^{r/<}(\omega)+ G_{BB}^{r/<}(\omega)}{2}
= \frac{G_{++}^{r/<}(\omega)+ G_{--}^{r/<}(\omega)}{2}.
\end{equation}
The energy distribution function $f(\omega)$ and the effective temperature $T_{\rm eff}$ are then defined exactly like in the case of the single-band metal, see the equations~(\ref{feff0}) and (\ref{distr0}).

\begin{figure}
\begin{center}
\rotatebox{0}{\resizebox{2.5in}{!}{\includegraphics{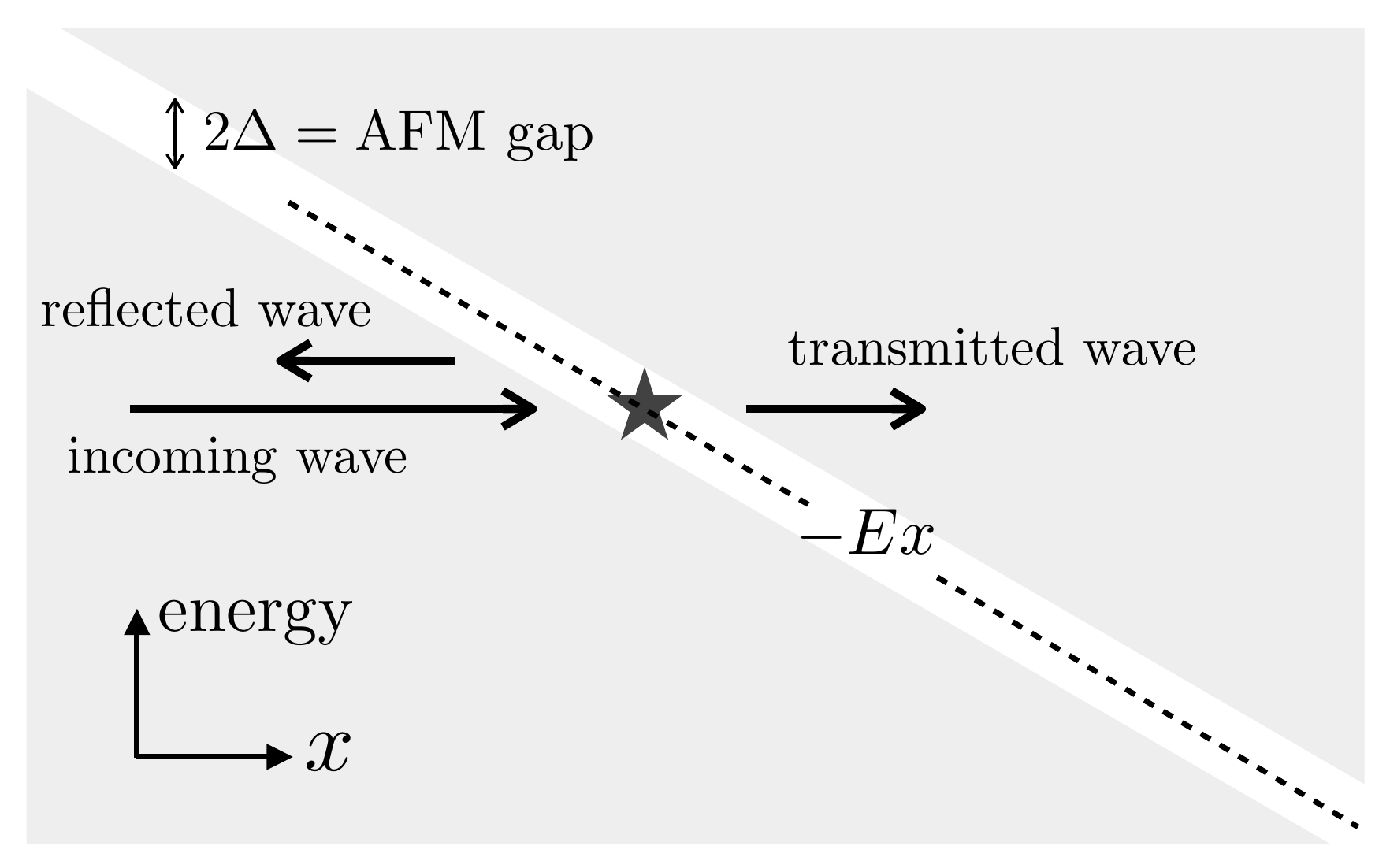}}}
\caption{Energy diagram of the anti-ferromagnet subject to a DC electric field. A staggered
order in lattice develops a bandgap given by the order parameter
$\Delta$. A uniform electric field $E$ is incorporated as a potential ramp
throughout the system. Thus, the gap acts as a potential barrier
for an incident wave approaching with the velocity $v_0$, which splits into a reflected and transmitted wave.
In the small-gap large-field limit, the amplitude of the transmitted
wave is proportional to the square-root of the Landau-Zener factor
$\rme^{-\pi\alpha/2}$ with $\alpha=\Delta^2/v_0E$.
}
\label{fig:scatt}
\end{center}
\end{figure}

\subsubsection{Analytic Solution}
We can solve for the eigen-function $\phi^\lambda_+$ by eliminating $\phi^\lambda_-$ in the coupled
equations~(\ref{diff1}) to obtain the second-order differential equation
\begin{equation} \label{eq:diff28}
v_0^2\partial_x^2\phi^\lambda_++[(\omega+Ex)^2-\rmi v_0E-\Delta^2]\phi^\lambda_+ = 0.
\end{equation}
Similarly to Zener's original paper~\cite{zener}, we use the
variables
\begin{equation}
z \equiv (2E/v_0)^{1/2}\rme^{\rmi\pi/4}(x+\omega/E)\mbox{ and } n \equiv \rmi\Delta^2/(2v_0E)-1
\end{equation}
to transform Eq.~(\ref{eq:diff28}) to the standard form
\begin{equation}
\frac{\rmd^2\phi^\lambda_+(z)}{\rmd z^2}+\left(n+\frac12-\frac14 z^2\right)\phi^\lambda_+(z) = 0.
\end{equation}
The solutions of this equation can be expressed in terms of the
parabolic cylinder function~\cite{whittaker,gradshteyn} as 
\begin{equation} \label{pm}
\phi^\lambda_+(x+\omega/E)\propto D_{-n-1}(\pm\rmi z),
\end{equation}
with $D_{-n-1}(\rmi z)$ decaying to
zero for $z \to \rme^{-\rmi\pi/4} \times \infty$ and $z \to \rme^{-\rmi(3\pi/4)}\times \infty$. 
We choose the sign in the argument of the parabolic cylinder function in Eq.~(\ref{pm})
according to the boundary condition of right- or left-incident
wavefunction. A more detailed discussion is given in Appendix~\ref{app:wf}.
The normalized solution for the right-moving wavefunction can be written
down as
\begin{equation}
\phi^R_+(x,\omega)  = 
\rme^{-3\pi\alpha/8}D_{-\rmi\alpha/2}\left(2y \, \rme^{-\rmi\pi/4} \right),
\label{phiR+}
\end{equation}
where we introduced the dimensionless parameters
\begin{equation}
\alpha \equiv \frac{\Delta^2}{v_0E}\mbox{ and }
y \equiv \sqrt{ \frac{E}{2v_0}}\left(x+\frac{\omega}{E}\right).
\label{alphay}
\end{equation}
Similarly, the eigen-function $\phi^\lambda_-(x,\omega)$ is given by
\begin{equation}
\phi^R_-(x,\omega)=-\sqrt{\frac{\alpha}{2}}\rme^{-\rmi\pi/4}\rme^{-3\pi\alpha/8}
D_{-\rmi\alpha/2-1}\left(2y \, \rme^{-\rmi\pi/4}\right).
\label{phiR-}
\end{equation}
The left-moving
solutions are obtained by symmetry:
\begin{eqnarray}
\phi^L_-(x,\omega) & = &
[\phi^R_+(-x,-\omega)]^*
=
\rme^{-3\pi\alpha/8}D_{\rmi\alpha/2}\left(- 2y \, \rme^{\rmi\pi/4} \right),
\nonumber \\
\phi^L_+(x,\omega) & = &
-[\phi^R_-(-x,-\omega)]^*.
\label{symmetry}
\end{eqnarray}

\section{Landau-Zener vs. In-Gap Tunneling Regimes}
We shall now distinguish two regimes: (i) the weakly gapped case, when  $\alpha = {\Delta^2}/{v_0 E} \ll 1$, for which the nonequilibrium excitations will be shown to be dominated by Landau-Zener tunneling events; (ii) the strongly gapped case, when  $\alpha  \gg 1$, dominated by the excitation of dissipative in-gap states.

\begin{figure}
\begin{center}
\rotatebox{0}{\resizebox{2.4in}{!}{\includegraphics{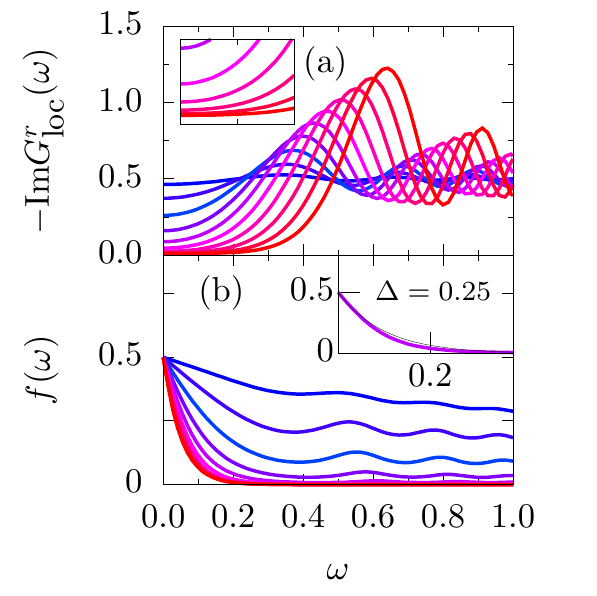}}}
\caption{
(a) Local density of states $-\mbox{Im }\,G^r_{\rm loc}(\omega)$ at
$E=0.05$, $\Gamma=0.01$, with $\Delta$ varying from $0.05$ (blue)
to 0.5 (red) in steps of 0.05. The energy unit is set by $v_0=1$. While the energy
gap rapidly develops with increasing $\Delta$, the inset (blown up near zero energy) reveals the presence of a finite density of in-gap states on the order of $\Gamma/(2\Delta v_0)$ (see text for details). (b) Distribution functions $f(\omega)$. The inset shows the agreement with the analytic result $f(\omega) \simeq \frac12 \rme^{-2\Delta\omega/v_0E}$ (black line) computed for large $\Delta$, already at $\Delta=0.25$.
}
\label{fig:spec}
\end{center}
\end{figure}

The distinct behaviors in these two regimes of (a) the local density of states and, (b) the energy distribution function, are illustrated in FIG.~\ref{fig:spec} which gives the numerical solutions computed at a fixed $E=0.05$ for increasing values of $\Delta$.
Quite naturally, the local density of states in FIG.~\ref{fig:spec}(a)
continuously develops an energy gap on the order $\Delta$.
In the small
gap (or large field) limit, the gap is filled up by the accelerating
electrons, as shown by the smearing of the gap.
Perhaps less obvious is the presence of a
small but finite density of in-gap states at  $|\omega|\lesssim\Delta$
in the large-gap, small-field limit. We shall see that they are due to the dissipation which, in our model, broadens the two bands and make them leak inside the gap.
They could also be due to the presence of impurities in the system.
The distribution functions in FIG.~\ref{fig:spec}(b) displays a rapid
crossover between a hot nonequilibrium steady state at small $\Delta$,
and a cooler state where excitations are localized in
$|\omega|\lesssim\Delta$ at large $\Delta$.

Below, we elucidate the different mechanisms at stakes, and their associated energy scales, by deriving the \emph{analytic} solutions of the nonequilibrium steady states in the two regimes.

\subsection{Landau-Zener Tunneling Regime}
In the small $\alpha = \Delta^2/v_0 E \ll 1 $ regime, an asymptotic expression of the wave function in Eq.~(\ref{phiR+}) can be worked out when $|y|\gtrsim\alpha$, reading
\begin{eqnarray}
\phi^R_+(y<0)  & \simeq & \rme^{\rmi y^2}|2y|^{-\rmi\frac{\alpha}{2}} 
+\frac{\sqrt{2\pi}\rme^{-\frac{\pi}{4}(\alpha-\rmi)}
}{\Gamma(\frac{\rmi\alpha}{2})}\rme^{-\rmi y^2}
|2y|^{\rmi\frac{\alpha}{2}-1}
\nonumber \\
\phi^R_+(y>0)  & \simeq & \rme^{-\frac{\pi\alpha}{2}}\rme^{\rmi y^2}
(2y)^{-\rmi\frac{\alpha}{2}},
\label{phiR}
\end{eqnarray}
where $\Gamma(x)$ is the Gamma function.
For convenience, let us focus on $\omega>0$.
The first term in the above expression of
$\phi^R_+(y<0)$ is the incident wave from $-\infty$. Indeed, it becomes the
free propagating wave computed in Eq.~(\ref{phi0}) in the limit $\alpha\to 0$. See also
FIG.~\ref{fig:scatt} for a brief discussion. The second term is the
reflected wave which is scattered from the gap. The term in $\phi^R_+(y>0)$ represents the transmitted
wave. For small $\alpha$, the amplitude of the reflected wave is small
with $| \sqrt{2\pi} \rme^{-\pi\alpha/4}/\Gamma(\rmi \alpha/2)|^2 \simeq  \pi \alpha^2/2$
and it can be neglected for  $|y|\gg 1$.
We may then approximate the right-moving wavefunction for all
$|y|\gtrsim\alpha$ by
\begin{equation}
\phi^R_+(x,\omega) \simeq \rme^{\rmi y^2}( 2y \, \rme^{-\rmi\pi})^{-\rmi\alpha/2}.
\end{equation}
$y$ is analytically continued to the complex plane with its phase restricted
to $0\leq{\rm arg}(y)<\pi$, with the branchcut on the negative real
axis. Then, on the positive real axis $y>0$, the factor
$(\rme^{-\rmi\pi})^{-\rmi\alpha/2}=\rme^{-\pi\alpha/2}$ gives the
Landau-Zener amplitude reduction while on the negative axis the factor is cancelled out. 

\begin{figure}
\begin{center}
\rotatebox{0}{\resizebox{3.0in}{!}{\includegraphics{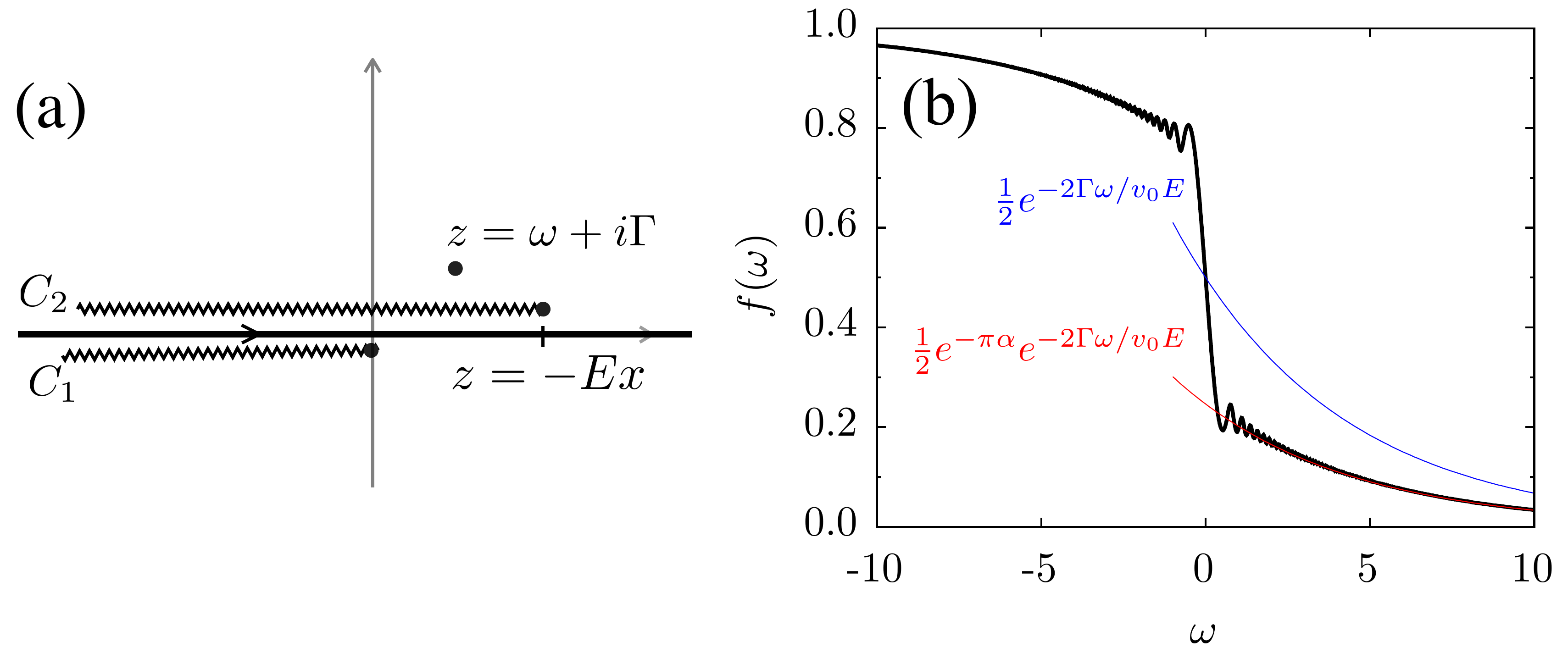}}}
\caption{\small (a) Singularities in the integral of the retarded GF, Eq.~(\ref{contourint}), in the Landau-Zener regime using the asymptotic expansion of the parabolic cylinder function. (b) Nonequilibrium distribution function $f(\omega)$ of the driven-dissipative AF computed numerically in the Landau-Zener tunneling (LZT) regime, and compared to the expression derived in Eq.~(\ref{eq:fw1}), showing a number of nonequilibrium excitations which is reduced by the factor $\rme^{-\pi\alpha}$ compared to the single-band metal.
}
\label{fig:contour}
\end{center}
\end{figure}

For $\omega>0$, and in the small-$\alpha$ limit, the right-moving wavefunctions traveling from $x<0$ contribute the most to the lesser GFs in Eq.~(\ref{lesserg}). We may therefore approximate the retarded GF by
\begin{equation}
G^r_{++}(0,x;\omega) = \int_{-\infty}^\infty\frac{\rho_{++}(0,x;\omega')}{
\omega-\omega'+\rmi\Gamma}\frac{\rmd\omega'}{2\pi v_0},
\label{contourint}
\end{equation}
with
\begin{align}
\rho_{++}(0,x;\omega')\simeq &
\rme^{-\pi\alpha}\rme^{-\rmi\frac{E}{2v_0}
(x^2+2x\omega'/E)}  \nonumber \\
& \times
\omega'^{-\rmi\alpha/2}(\omega'+Ex)^{\rmi\alpha/2}.
\end{align}
As represented in FIG.~\ref{fig:contour}(a), there are two branchcuts: $C_1\equiv (-\rmi\epsilon,
-\infty-\rmi\epsilon)$ for $\omega'^{\rmi\alpha/2}$ and $C_2\equiv(-x+\rmi\epsilon,
-\infty+\rmi\epsilon)$ for $(\omega'+Ex)^{-\rmi\alpha/2}$.
This choice of
branchcuts ensures that the complex power functions coincide with the
integrand everywhere on the real axis. The main contribution to the integral
is the residue at $\omega'=\omega+\rmi\Gamma$,
and we detail its computation in the
Appendix~\ref{app:intLZ}. The resulting retarded GF for $x\lesssim
-(2v_0/E)^{1/2}\alpha$ is
approximately
\begin{eqnarray}
G^r_{++}(0,x;\omega) & \simeq &
\frac{-\rmi}{v_0}\rme^{-\frac{\rmi Ex^2}{2v_0}-\frac{(\Gamma-\rmi\omega)|x|}{v_0}}
\nonumber \\
& & \times  \left|\frac{\omega+\rmi\Gamma}{\omega-E|x|+\rmi\Gamma}
\right|^{-\rmi\alpha/2}\rme^{-\alpha\varphi/2},
\label{finalgr}
\end{eqnarray}
where
\begin{equation}
\varphi = \pi
-\tan^{-1}\frac{\Gamma}{\omega}-\tan^{-1}\frac{\Gamma}{E|x|-\omega}.
\end{equation}
Note that Eq.~(\ref{finalgr}) is invalid at $x=0$ and therefore cannot be used to compute $G^r_{\rm loc}(\omega)$. As shown in FIG.~\ref{fig:spec}(a) for small $\alpha$, the spectral
function approaches the simple-metal limit, $-{\rm Im}G^r_{\rm
loc}(\omega>\Delta)\approx 1/(2v_0)$, away from the gap.

\begin{figure}
\begin{center}
\rotatebox{0}{\resizebox{3.5in}{!}{\includegraphics{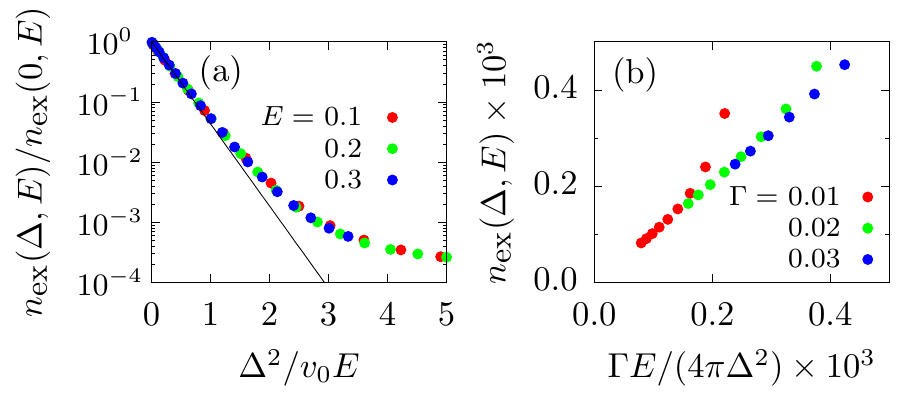}}}
\caption{Number of electronic excitations $n_{\rm ex}(\Delta,E)$.  (a) In the small-gap (or large-field) regime, it follows the Landau-Zener theory (black line).
(b) In the large-gap (or small-field) regime, the excitations are limited
inside the gap with $n_{\rm ex}(\Delta,E)=\Gamma E/(4\pi \Delta^2)$.
}
\label{fig:nex}
\end{center}
\end{figure}

In the small damping limit, we get $\varphi\simeq \pi$, and the local lesser GF
can be approximated as
\begin{equation}
G^<_{\rm loc}(\omega > 0) \simeq \frac{\rmi}{v_0}\,  f(\omega).
\end{equation}
The energy distribution function then becomes
\begin{equation} \label{eq:fw1}
 f(\omega > 0) \simeq \frac12  \rme^{-\pi\alpha} \rme^{-2\Gamma\omega/v_0E}.
\end{equation}
The numerical effective distribution in FIG.~\ref{fig:contour}(b) shows
an excellent agreement with the analytic result. The step-like drop of
$f(\omega)$ near $\omega=0$ by the Landau-Zener factor
demonstrates a clear departure from a thermal distribution and highlights
the electronic nature of the population inversion.
Compared to the driven-dissipative single-band metal studied in Section~\ref{sec:metal}, see Eq.~(\ref{eq:glessloc}),  the number of excited sates is reduced by the Landau-Zener factor $\rme^{-\pi\alpha}$.
This is reflected in the effective temperature
\begin{align}
T_{\rm eff}(\alpha \ll 1)  =  \sqrt{\frac32}\frac{v_0 E }{\pi \Gamma} \rme^{-\pi\alpha},
\end{align}
which is also reduced compared to the single-band metal in Eq.~(\ref{eq:teff0}),
as well as in the numerical calculations in FIG.~\ref{fig:nex}(a) which displays the total number of excitations
\begin{equation}
n_{\rm ex}(\Delta,E)=2 \int_{0}^D G^<_{\rm loc}(\omega)\frac{\rmd\omega}{2\pi\rmi}.
\end{equation}
The excitation density $n_{\rm ex}(\Delta,E)$ is defined
as the electron and hole excitations from the zero-field electron
distribution.
Here, we introduced an energy cutoff $D$ (set to $10\, v_0$ throughout the
paper) to regularize the linear dispersion relation.
For small $\alpha$, the agreement with the Landau-Zener factor (black
line) is excellent, as also previously demonstrated in the lattice model
calculation~\cite{nanolett}.

Note that in the regime $\alpha \to 0$, the distribution function in Eq.~(\ref{eq:fw1}) naturally boils down to the one of the single-band metal in Section~\ref{sec:metal}, see Eq.~(\ref{distr0}). 

\subsection{In-Gap Tunneling Regime}

In the opposite regime of large $\alpha = \Delta^2 / v_0  E \gg 1$,
\textit{i.e.} with a large gap or a small field, the electronic
transport proceeds quite differently. This is illustrated in
FIG.~\ref{fig:nex}(a) which shows a strong deviation of the  total
number of excitations from the Landau-Zener theory.  The spectral weight
inside the gap is now controlled by the dissipation,
bounded from below by the zero-field spectral weight
$-\mbox{Im }G^r_{\rm loc}(0)=\Gamma/(2v_0 \Delta)$. The spectral
properties deep inside the gap can be approximated by the zero-field
retarded GF, as detailed in Appendix~\ref{sec:zerofield}.  The
electronic excitations are most efficient within the gap, as
demonstrated by the energy distribution function $f(\omega)$ displayed
in FIG.~\ref{fig:spec}(b).  The shape of the distribution function in
this regime can be understood as follows.  In the case of the
single-band metal studied in Section~\ref{sec:metal}, the damping rate
$\Gamma$ was controlling the energy window of the nonequilibrium
excitations. In the presence of a gap, the gap acts as a potential
barrier and provides a decay rate similar to the WKB theory.  Therefore, the
gap parameter $\Delta$ replaces $\Gamma$ in Eq.~(\ref{distr0}), leading
to the energy distribution function
\begin{equation} 
f(\omega > 0)\simeq \frac12 \rme^{-2\Delta\omega/v_0 E}
\label{feff1}
\end{equation}
for large $\alpha$.  
It is interesting to note that while dissipation is
essential to create the in-gap states, the distribution function
has negligible dependence on the damping parameter $\Gamma$.
Noteworthy enough, while FIG.~\ref{fig:nex}(a) indicates that the
deviation from the Landau-Zener theory starts
around $\alpha\sim 1-2$, the inset of FIG.~\ref{fig:spec}(b) shows that the distribution function in Eq.~(\ref{feff1}) computed for $\alpha \gg 1$ is already valid at  $\Delta=0.25$.

The corresponding effective temperature can be computed as
\begin{align} \label{eq:teff1}
T_{\rm eff}(\alpha \gg 1) = \sqrt{\frac{3}{2}} \frac{v_0 E}{\pi |\Delta|},
\end{align}
that is much smaller than in the LZ regime: $T_{\rm eff}(\alpha \gg 1) \ll T_{\rm eff}(\alpha \ll 1)$.
The total number of nonequilibrium charge excitations is then well approximated by
\begin{equation}
n_{\rm ex}(\Delta,E)\simeq 2\int_0^\infty \Ud{\omega} \frac{\Gamma}{2\pi
v_0\Delta}f(\omega)=\frac{\Gamma E}{4\pi \Delta^2},
\end{equation}
which is confirmed by the numerical calculations presented in
FIG.~\ref{fig:nex}(b).

Note that the effective temperature above is seemingly independent of
$\Gamma$. Indeed, it was computed in the regime $|\omega| \leq \Delta$
where the energy distribution is given by the expression in
Eq.~(\ref{feff1}). At large frequencies $\omega\gg\Delta$, however,
$f(\omega)$ is expected to behave as $\frac12
\rme^{-\pi\alpha}\rme^{-2\Gamma\omega/v_0E}$.  This tail contributes a
correction term proportional to $\rme^{-\pi\alpha}(\Delta/\Gamma)^2$,
which grows large in the $\Gamma\to 0$ limit. This is consistent with
the previous works~\cite{eckstein2010,eckstein2011} in which the
effective temperature has been shown to diverge in dissipationless
driven systems.  In the following Section, however, we limit the energy
integrals at the cutoff energy $D$ and, furthermore, the decaying
integrand in the gap equation renders insignificant the effect of the
$f(\omega)$ tail, particularly in the large $\alpha$ limit as shown in
the inset of FIG.~\ref{fig:spec}(b).  


\section{Mean-Field Theory of Resistive Switching in Anti-Ferromagnets}

In the previous Section, we discussed how, upon increasing the electric
field and keeping the gap parameter $\Delta$ fixed, the electrons are
initially excited via in-gap tunneling events, and then undergo
Landau-Zener tunneling processes as the field is further increased.  In
a recent paper by the Authors~\cite{nanolett}, it has been shown that an
inhomogeneous mean-field (MF) approach on a two-dimensional Hubbard
model could capture the hysteretic nature of the true resistive
switching transition: sweeping up and down the voltage bias applied on
a finite-size two-dimensional lattice resulted in an
insulator-to-metal transition (IMT) and a metal-to-insulator transition
(MIT), separated by a region of bi-stability.  Importantly, this
nonequilibrium bi-stability was found to be crucial to explain the
abrupt nature of the resistive switching, independently whether the
parent equilibrium transition is continuous or discontinuous.

Here, we explore the mechanisms behind the IMT and the MIT via a
continuum theory which allows an \emph{analytic} understanding of the
problem. Below, we develop the mean-field theory for a
driven-dissipative anti-ferromagnet (AF). This approach may be extended
to other types of order without much difficulty.  We start with the
standard single-orbital Hubbard model, with on-site repulsive Coulombic
interaction 
$\hat{V}=U\sum_i(\hat{n}_{i\uparrow}-\bar{n})(\hat{n}_{i\downarrow}
-\bar{n})$
with the electron number operator
$\hat{n}_{i\sigma}=d^\dagger_{i\sigma}d_{i\sigma}$, the Coulomb
parameter $U$, and the on-site occupation expectation value
(averaged over spin)
$\bar{n}$. 
The emergence of an AF phase corresponds to the breaking of the
translational invariance of the lattice into a staggered order. The
energy levels of the two resulting sublattices $A$ and $B$ get shifted
alternately by $\pm \Delta$, the AF order parameter which opens a charge gap.
The corresponding mean-field decoupling of the Hubbard interaction consists in replacing
\begin{equation}
\hat{V} \mapsto \hat{V}_{\rm MF}=\Delta
\sum_{m=A/B,i_m,\sigma}(-1)^m\sigma \hat{n}_{i_m,\sigma},
\end{equation}
with the sublattice index $m$ and $(-1)^m=\pm 1$ for $m=A$ and $B$,
respectively. $\hat{n}_{i_m}$ is the electron occupation on the $i_m$-th
site within the $m$-sublattice. The resulting theory is invariant under $\Delta \mapsto -\Delta$, and we may work with $\Delta \geq 0$.
Since the MF Hamiltonian is diagonal in the spins, we may also afford to ignore the spin degrees of freedom in what follows.
The nonequilibrium self-consistent equation on the AF order parameter, often referred as the gap equation, reads
\begin{align}
\Delta &= \frac{U}{2}(\langle
n_A\rangle-\langle n_B\rangle) \\
& = 
U\int_{-D}^D\left[G^<_{-+}(\omega)+G^<_{+-}(\omega)\right]\frac{\rmd\omega}{2\pi\rmi}.
\label{gapeq}
\end{align}

\subsection{Equilibrium Phase Transition}
For reference, let us briefly review the conditions for the equilibrium, temperature-driven, phase
transition. The mean-field approach predicts a second
order phase transition~\cite{slater}.
As described in Appendix~\ref{sec:zerofield}, 
for $|\omega|>\Delta_0$ and at zero temperature, the gap equation~(\ref{gapeq}) becomes
\begin{equation}
\frac{2\pi
v_0}{U}=\int_{-D}^{-\Delta_0}\frac{\rmd\omega}{\sqrt{\omega^2-\Delta_0^2}}
\simeq \ln\left(\frac{2D}{\Delta_0}\right),
\end{equation}
in the small gap limit, $\Delta_0\ll D$. This yields the familiar
expression for the order parameter at zero temperature and zero-field
\begin{equation}
\Delta_0\simeq 2D\exp\left(-\frac{2\pi v_0}{U}\right).
\label{delta0}
\end{equation}
The transition temperature $T_{\rm N}$ is set by the finite-temperature gap equation
\begin{equation}
\frac{2\pi v_0}{U} = 
-\int_{-D}^D\frac{\omega f_0(\omega)}{\omega^2+\Gamma^2}\rmd\omega
\simeq
\ln\left(\frac{2D}{\pi T_{\rm N}}\right)+\gamma,
\end{equation}
where $ f_0(\omega)$ is the Fermi-Dirac distribution at the temperature $T_{N}$ and $\gamma\approx 0.577$ is the Euler constant.
This allows to relate the  N\'eel temperature to the zero-temperature gap via
\begin{equation}
T_{\rm N}\approx \frac{\rme^\gamma}{\pi}\Delta_0\approx 0.57\,\Delta_0.
\label{tn}
\end{equation}

\begin{figure}
\begin{center}
\rotatebox{0}{\resizebox{3.0in}{!}{\includegraphics{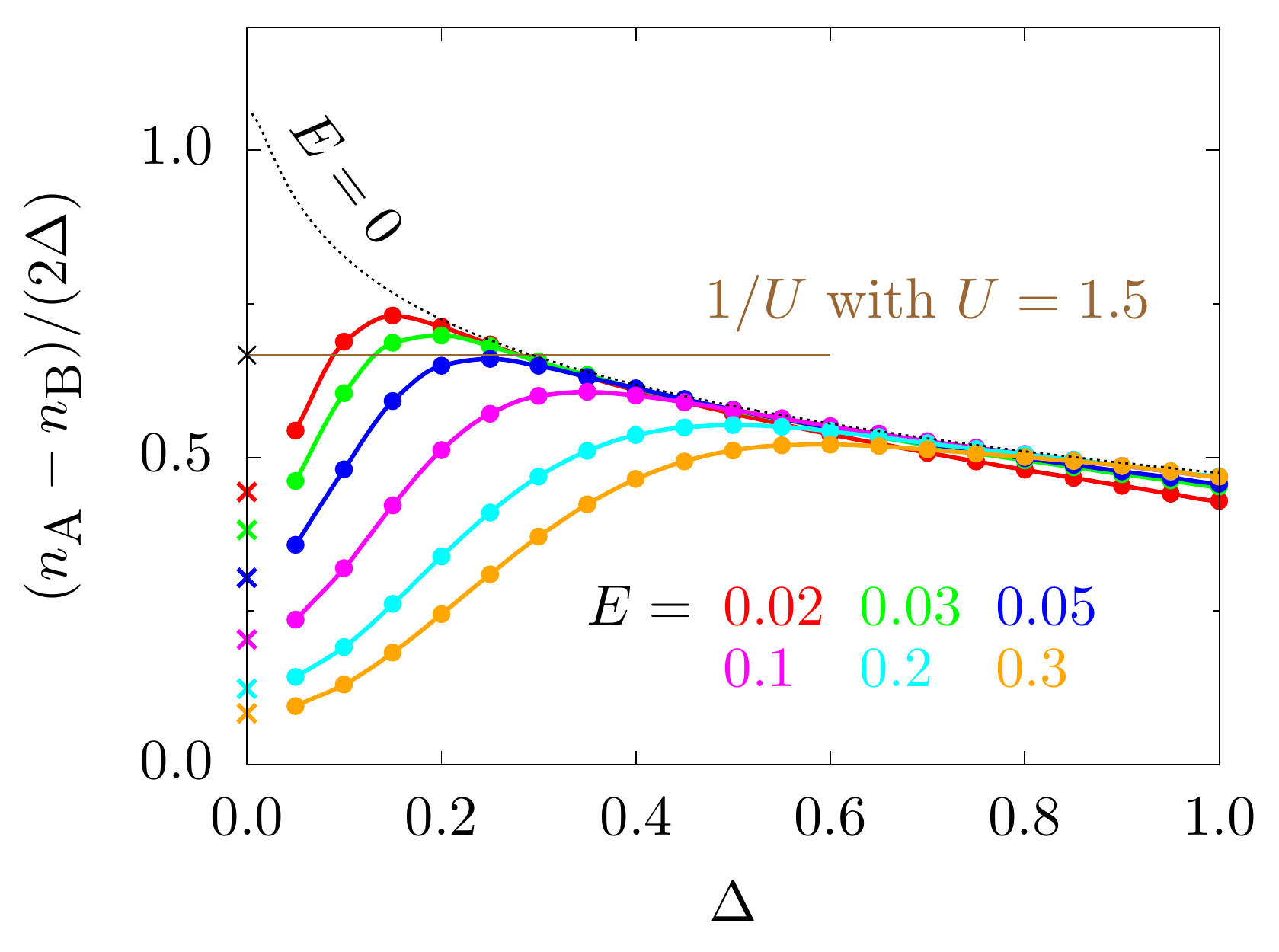}}}
\caption{Nonequilibrium mean-field self-consistent condition on the order parameter $\Delta$. The solutions correspond to the intersection of the curves at different $E$-field with the $1/U$ line. 
The finite-$\Delta$ solutions with a negative slope are the stable solutions.
The solution at $\Delta\approx 0.25$ brutally disappears at $E_{\rm MIT}=0.044$. 
The crosses at $\Delta = 0$ are calculated as discussed in Appendix~\ref{sec:mit}.
}
\label{fig:gap}
\end{center}
\end{figure}

\subsection{Nonequilibrium Phase Transitions} \label{sec:Neq}

\subsubsection{Numerical Results}
The numerical solutions of the nonequilibrium mean-field self-consistent gap equation are
presented in FIG.~\ref{fig:gap}, where the ({\sc rhs})$/(U\Delta)$ of
Eq.~(\ref{gapeq}) is plotted as a function of the ({\sc lhs}).
$\Delta=0$ is always a trivial solution, marked by crosses
in the figure. It corresponds to an ungapped, metallic, phase.
The intersections of the curves with $1/U$ at finite values of $\Delta$ are  
non-trivial solutions that correspond to anti-ferromagnetic states.
In equilibrium ($E=0$), the curve (dotted line) is monotonic in $\Delta$ and thus supports a single AF solution. As $U$ is varied, the single-valued
order parameter $\Delta$ evolves continuously from a vanishing to a finite value. This is the second order equilibrium phase transition.
However, as $E$ is turned on, the curve becomes non-monotonic and
allows two AF solutions  at small enough $E$. A stability analysis indicates that the solution with the smaller $\Delta$ is unstable while the other one is stable. When $E$ becomes larger than a critical value ($E_{\rm MIT}=0.25$ in the
figure with $U = 1.5$), the two AF solutions suddenly disappear, leaving $\Delta=0$ as the only solution. This is the IMT: a strongly discontinuous nonequilibrium phase transition which emerges out of a continuous transition in equilibrium~\cite{nanolett}.

Below, we discuss the quantitative criteria for the switching electric fields at the IMT and MIT.

\subsubsection{Insulator-to-Metal Transition}
As discussed above, the IMT occurs when the stable finite-$\Delta$ solution abruptly ceases to exist, at the field $E_{\rm IMT}$.
We first determine in which of the regimes, Landau-Zener or
in-gap tunneling, the IMT occurs. In FIG.~\ref{fig:gap}, the IMT occurs at
$\alpha_{\rm IMT}\approx 1.25$, thus in the
crossover region between the two limiting regimes. However, as shown in the inset
of FIG.~\ref{fig:spec}(b), the distribution function in Eq.~(\ref{feff1}), although computed far in the in-gap tunneling regime (\textit{i.e.} for $\alpha \gg 1$),
describes the numerical solution fairly well. We use it, together with the approximation that the off-diagonal components of the GFs can be replaced by their equilibrium  components (see Appendix~\ref{sec:goff}), to re-write the gap equation as
\begin{align}
\frac{2\pi v_0}{U} & = 
\int_{-D}^{-\Delta}\frac{\rmd\omega}{\sqrt{\omega^2-\Delta^2}}
-\int_\Delta^\infty
\frac{\rme^{-2\Delta\omega/v_0E}}{\sqrt{\omega^2-\Delta^2}}\rmd\omega, \label{gapimt0}
\end{align}
or, equivalently,
\begin{align}
0 & =  \ln\left(\frac{\Delta}{\Delta_0}\right)
+K_0\left(\frac{2\Delta^2}{v_0E}\right).
\label{gapimt}
\end{align}
See the derivation of Eq.~(\ref{gapapp}) in Appendix for more details.
$K_n(x)$ is the modified Bessel function of the second kind.
The first term of the ({\sc rhs}) in Eqs.~(\ref{gapimt0}) and~(\ref{gapimt})
is the equilibrium contribution, and the second term is the reduction of
the order parameter due to the nonequilibrium excitations where the
main contribution originates from the edge of the gap $\omega\simeq\Delta$.

\paragraph*{Threshold field.}
The condition for the IMT is that the derivative of the {\sc rhs} of the
above equation with respect to $\Delta$ vanishes at the solution, \textit{i.e.}
$1+4\alpha K_0'(2\alpha)=0$. This yields $\alpha_{\rm IMT}\approx 0.63$. Note the relative discrepancy with
the numerical result given above, $\alpha_{\rm IMT}\approx 1.25$.
It shows that the analytic
derivation underestimates $\Delta_{\rm IMT}$ and overestimates $E_{\rm
IMT}$, due the piece of integral that was neglected inside the gap.
Substituting the analytic result $\alpha_{\rm IMT}\approx 0.63$ into
Eq.~(\ref{gapimt}), we obtain
\begin{equation}
\Delta_{\rm IMT}\simeq \rme^{-K_0(2 \alpha_{\rm IMT})}\Delta_0
\approx 0.74\,\Delta_0,
\end{equation}
and
\begin{equation}
E_{\rm IMT}=\frac{\Delta_{\rm IMT}^2}{\alpha_{\rm IMT} v_0}\approx
0.88\,\frac{\Delta_0^2}{v_0}.
\label{eq:eimt}
\end{equation}
From the numerical calculations with $U=1.5$, we have $\Delta_0=0.30$,
$E_{\rm IMT}=0.044$
and $\Delta_{\rm IMT}=0.23$, yielding the ratios
$\Delta_{\rm IMT}\approx 0.76\,\Delta_0$ and $E_{\rm IMT}\approx
0.49\,(\Delta_0^2/v_0)$, which are in a reasonable agreement with the analytic estimates.

Importantly, these results elucidate a long standing problem: the
puzzling small values of the electric field that are needed to achieve
the IMT. Indeed, our solution shows that 
\begin{align}
E_{\rm IMT}/\Delta_0\sim \Delta_0/v_0\ll 1,
\end{align}
\textit{i.e.} that the energy scale of the switching field can be up to
one order of magnitude smaller than the energy gap. However, with a
typical $\Delta_0 \sim 0.1$~eV and $\hbar v_0/a \sim 1$~eV, this
corresponds to switching fields on the order of $E_{\rm IMT} \sim
10^2$~kV/cm which are still one to two orders of magnitude larger than
what is observed experimentally. We have seen in the previous
work~\cite{nanolett} that nucleation of conducting filament in spatially
inhomogeneous systems reduces $E_{\rm IMT}$ significantly. We shall also
argue in Sect.~\ref{sec:finiteT} that the remaining discrepancy can be
much reduced by working with an environment at a finite temperature
$T_{\rm b} \lesssim T_{\rm N} $ rather than $T_{\rm b} =0$,
\textit{i.e.} by bringing the equilibrium system closer to its N\'eel
transition, which is the case in most experiments.


\paragraph*{Effective temperature at the transition.}
Another crucial test for the theory is the ability to predict that the effective
electronic temperature at the IMT matches the
equilibrium transition temperature $T_{\rm N}$, as it has recently been
demonstrated experimentally~\cite{zimmers}. From Eqs.~(\ref{teff}) and
(\ref{feff1}), we obtain the analytic estimate
\begin{equation} 
T_{\rm IMT}=\frac{\sqrt{6}}{2\pi}\frac{v_0 E_{\rm IMT}}{\Delta_{\rm IMT}}
\approx 0.46\,\Delta_0\approx 0.81\times T_{\rm N}.
\label{timt}
\end{equation}
The numerical results give $T_{\rm IMT}=0.163=0.54\,\Delta_0\approx 0.95\times T_{\rm N}$, in very good agreement with the previous numerical work~\cite{nanolett} on discrete lattices. 
This proves that the effective temperature at which the IMT occurs is simply controlled by $T_{\rm N}$, the equilibrium transition temperature. This is one of the main result of this work, which justifies recent experimental observations made in Ref.~\cite{zimmers}.

The IMT condition can be roughly understood as the situation when the tail of the
electron distribution in Eq.~(\ref{feff1}) begins to overlap
with density of states at the edge of the gap, $v_0E/(2\Delta)\sim \Delta$
\textit{i.e.} $ \alpha \sim 0.5$, and the number of nonequilibrium excitations is about to
proliferate.
It is remarkable that, despite the IMT occurring in the crossover region
between the Landau-Zener and the in-gap tunneling regimes, the IMT conditions do
not depend sensitively on the dissipation parameter $\Gamma$. This
clearly indicates that the IMT is fundamentally an electronic process,
while it also permits a thermal interpretation. 

\paragraph*{$I-V$ scaling near the IMT.}
Based on the gap equation in Eq.~(\ref{gapimt}), we can analyze the
limiting behavior near the IMT. Writing $\Delta=\Delta_{\rm
IMT}+\delta\Delta$ and $E=E_{\rm IMT}+\delta E$, and expanding the gap
equation to the lowest orders, we obtain
\begin{equation}
0 = 1.86\,\frac{(\delta \Delta)^2}{2\Delta_{\rm IMT}^2}
+\frac{\delta E}{2E_{\rm IMT}},
\end{equation}
which can be massaged to the typical MF scaling relation
\begin{equation}
\delta \Delta\simeq \alpha_{\rm IMT}(-v_0\,\delta E)^{1/2} \mbox{ for }\delta \Delta>0,\,\delta E<0.
\end{equation}
Furthermore, the GFs do not have any singularities when $\Delta$
and $E$ pass through $\Delta_{\rm IMT}$ and $E_{\rm IMT}$, as can be
seen in FIG.~\ref{fig:spec}. Therefore, we may expand the
electric current $J$ around its value right before the IMT in powers of
$\delta \Delta$ and $\delta E$,
\begin{eqnarray}
J(E,\Delta) & \simeq & J_{\rm IMT}+a_E\,\delta E+a_\Delta\,\delta\Delta \nonumber \\
& \simeq & J_{\rm IMT}-\alpha_{\rm IMT}|a_\Delta|(-v_0\,\delta E)^{1/2},
\end{eqnarray}
where $a_E$ and $a_\Delta$ are expansion coefficients. Since the current
is reduced when the gap increases, we must have $a_\Delta<0$. 
While the precise value of the above critical exponent in the current
characteristic is the result of a mean-field approach, and might
therefore get final-dimensional corrections, such a non-analytic and
rapid increase of the current close to the IMT is a universal prediction
of the theory. As a matter of fact, it has already been observed in our
previous numerical lattice simulation~\cite{nanolett} and in recent experiments~\cite{zimmers,doron}, and deserves closer scrutiny.

\subsubsection{Metal-to-Insulator Transition}
\paragraph*{Threshold field.}
The MIT is determined by the loss of stability of
the $\Delta=0$ solution. 
In FIG.~\ref{fig:gap}, the threshold $E_{\rm MIT}$ corresponds to when the curve at $\Delta=0$ (black cross) matches
$1/U$.
The stability of the metal is given by the condition
\begin{equation}
\frac{1}{U}\geq \lim_{\Delta\to
0}\frac{1}{2\Delta}(n_A-n_B).
\label{mitcrit}
\end{equation}
We can easily and accurately pinpoint the MIT by using perturbation theory in the small
$\Delta$ limit. The details are given in Appendix~\ref{sec:mit}. Equation~(\ref{mitcrit}) can be estimated as
\begin{equation}
\frac{1}{U}\gtrsim \frac{1}{2\pi v_0}\ln\left(\frac{2\gamma\Gamma D}{v_0
E}\right),
\label{mitcrit2}
\end{equation}
which typically overestimates the exact numerical integrals by less than
5\%. It yields a switching field 
\begin{equation}
E_{\rm MIT}=1.78\,\frac{\Gamma}{v_0}\Delta_0.
\label{emit}
\end{equation}
From the numerical calculations with $U=1.5$ and $\Gamma = 0.01$, we obtained $E_{\rm MIT}=0.0044$ yielding the ratio $E_{\rm MIT}=1.46\,(\Gamma / v_0 )\Delta_0$ which is, again, in good agreement with the analytic estimate.
Equation~(\ref{emit}) reveals that, unlike the IMT, the MIT crucially depends on the dissipation,
which is not surprising since the transition is initiated from a
metallic phase where Joule heating is non-negligible.
Most importantly, this also shows that $E_{\rm MIT}\sim(\Gamma/\Delta_0)E_{\rm IMT}$, therefore predicting the hierarchy
\begin{align}
E_{\rm MIT} \ll E_{\rm IMT} \ll \Delta_0.
\end{align}

\paragraph*{Effective temperature at the transition.}
Coming from a metallic regime, the effective temperature corresponds to the one computed in Eq.~(\ref{eq:teff0}). We obtain the following analytic estimate 
\begin{align} 
T_{\rm MIT} \approx 1.22 \times T_{\rm N},
\label{tmittn}
\end{align}
where $T_{\rm N}$ is the equilibrium N\'eel temperature. Once again,
this validates the idea that the resistive transitions can be
interpreted in the language of thermal transitions where the temperature
is replaced by an effective temperature accounting for the number of
excitations above the chemical potential. Noteworthy, while our
homogeneous mean-field approach cannot capture it, the bi-stability
region has already been shown to support the formation of metallic
filaments (and insulating domains) at lower effective
temperatures~\cite{nanolett}, \textit{i.e.} at lower threshold fields
than the above mean-field prediction~(\ref{tmittn}).

\subsection{RS at Finite Bath Temperature}\label{sec:finiteT} 

So far, we
have limited our theoretical analysis to the case of a zero-temperature
bath, $T_{\rm b} = 0$, and found switching fields one to two orders of
magnitude larger than what is typically observed in experiments (see the
discussion in Sect.~\ref{sec:Neq}). However, the experimental
measurements of the RS are often conducted close to room temperature.
Indeed, at low temperature the switching fields tend to be fairly large,
which is difficult to realize and can damage the samples.  Moreover, two
experimental observations are worth mentioning. First, as reported
previously~\cite{tlwu,sujay}, the switching fields $E_{\rm IMT}$ and
$E_{\rm MIT}$ show a significant temperature dependence, for instance
with $E_{\rm IMT}$ varying by a factor of two over 30 K near the N\'eel
temperature in VO$_2$~\cite{sujay}.  Second, it has recently been
reported in the superconductor-insulator switching~\cite{doron} that the
nonequilibrium phase transition displays critical behaviors similar to
the equilibrium liquid-gas transition close to its critical temperature,
with a strong temperature dependence of the switching electric field.
All these considerations lead us to investigate the case of a
finite-temperature bath, $T_{\rm b} > 0$. We shall show that increasing
$T_{\rm b}$ naturally corresponds to a higher effective temperature
$T_{\rm eff}(T_{\rm b})$, bringing the system closer to its transition,
therefore reducing the threshold fields and the intensity of the
nonequilibrium effects. As the bath temperature reaches the N\'eel
temperature,  $T_{\rm b} \to T_{\rm N}$, the threshold fields must
vanish, $E_{\rm IMT/IMT} \to 0$, and the nonequilibrium RS is expected
to progressively evolve into the continuous Ising transition of the
equilibrium N\'eel transition.

\subsubsection{Single-Band Metal}
We first discuss the impact of a finite temperature of the environment in the case of the single-band metal
studied in Section~\ref{sec:metal}. 
The equation~(\ref{lesser0}) is generalized to 
\begin{equation}
G^<_{\rm loc}(\omega)
=2\rmi\Gamma \int_{-\infty}^{\infty} \Ud{x} \frac12\sum_\lambda|G^r_\lambda(0,x;\omega)|^2 f_0(\omega + Ex),
\label{lesserT}
\end{equation}
where $f_0(\omega)$ is the Fermi-Dirac distribution function at temperature $T_{\rm b}$.
Using the definition of the effective temperature in Eq.~(\ref{teff}) and 
 the following identity~\cite{kirillov}
\begin{equation}
\int_0^\infty\omega[f_0(\omega+Ex)+f_0(\omega-Ex)]\rmd\omega
=\frac12 (Ex)^2+\frac16(\pi T_{\rm b})^2
\end{equation}
for an arbitrary bath temperature $T_{\rm b}$, we obtain the following effective temperature for the electric-field driven one-dimensional electron
gas coupled to a finite-temperature bath
\begin{equation}
T_{\rm eff}(T_{\rm b})^2=\frac32\left(\frac{v_0E}{\pi\Gamma}\right)^2+T_{\rm b}^2
=T_{\rm eff}(0)^2+T_{\rm b}^2.
\label{tefftb}
\end{equation}
Note that this relation can also be obtained by using the energy balance
between the Joule heating and the heat dissipation that compensate each other in the steady state~\cite{altshuler}.

\subsubsection{Finite-Temperature IMT}

\begin{figure}
\begin{center}
\rotatebox{0}{\resizebox{3.0in}{!}{\includegraphics{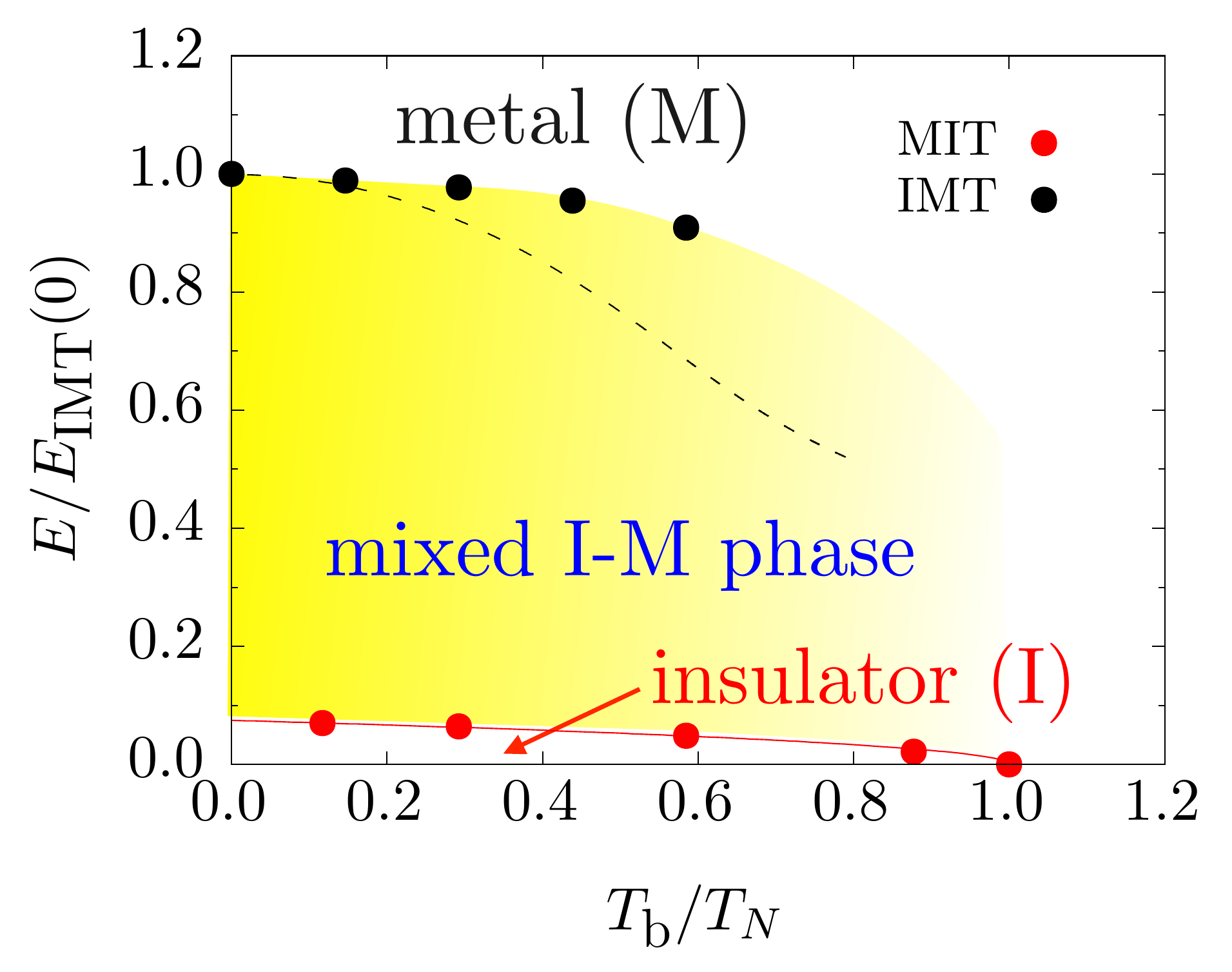}}}
\caption{Phase diagram in (electric field)-(bath temperature) space. Numerically computed switching
fields are shown as circles at the bath
temperature $T_{\rm b}$ for the IMT (black) and MIT (red). 
The dashed line is the analytic result Eq.~(\ref{parametric}), 
and the red solid line Eq.~(\ref{mitt}). $T_{\rm N}$ is the N\'eel temperature
at equilibrium, and
$E_{\rm IMT}(0) \approx 0.044$ the numerically estimated value at zero bath
temperature.
}
\label{fig:evst}
\end{center}
\end{figure}

We now turn to the case of the driven-dissipative anti-ferromagnet.
To compute the temperature dependence of the threshold field, $E_{\rm IMT}(T_{\rm b})$, we use perturbation theory in $T_{\rm b}$ around the previous results obtained at  $T_{\rm b} =0$.  Using the small-field approximations developed in Appendix~\ref{sec:zerofield}, we can generalize the gap equation in Eq.~(\ref{gapimt}) to 
\begin{equation}
0 = \ln\left(\frac{\Delta}{\Delta_0}\right)+
\left[1+\frac23\left(\frac{\pi T_{\rm b}\Delta}{v_0E}\right)^2\right]
K_0\left(\frac{2\Delta^2}{v_0E}\right).
\end{equation}
where we assumed $T_{\rm b} \ll T_{\rm eff}(0) \sim v_0 E / \Delta$.
Using the same criteria as Section~\ref{sec:Neq}, the IMT
can be parametrically solved as
\begin{eqnarray}
\frac{v_0E_{\rm IMT}}{\Delta_0^2} & = &
\frac{2}{u}\rme^{-h(u)},\nonumber \\
\frac{\pi T_{\rm b}}{\Delta_0} & = &
\left(\frac{3(1-2uK_1)}{u^2(uK_1-K_0)}\right)^{1/2}\rme^{-h(u)/2},
\label{parametric}
\end{eqnarray} 
with $K_0=K_0(u)$, $K_1=K_1(u)$ and $h(u)=(1-2K_0)K_0/(uK_1-K_0)$.
The solution is plotted with a black dashed line in FIG.~\ref{fig:evst}. By
expanding the relations around  $T_{\rm b}=0$, we
obtain the following expression for the $E_{\rm IMT}$,
\begin{equation} \label{eq:EIMTTb}
E_{\rm IMT}(T_{\rm b})\simeq E_{\rm IMT}(0)\left[1-0.88\left(\frac{T_{\rm
b}}{T_{\rm N}}\right)^2\right]
\mbox{ for } T_{\rm b}\ll T_{\rm N}.
\end{equation}
A numerical evaluation of $E_{\rm IMT}(T_{\rm b})$, represented by black circles in FIG.~\ref{fig:evst}, confirms its relatively slow decrease as the bath temperature is increased. 
At higher temperatures, $T_{\rm b} \sim T_{\rm N}$, the parametric solution in Eq.~(\ref{eq:EIMTTb}) ceases to be valid, and the numerical calculations are very hard to converge,  preventing us to resolve how $E_{\rm IMT}(T_{\rm b})$ approaches 0 when $T_{\rm b} \to T_{\rm N}$.
However, measurements in Ref.~\cite{tlwu} reported that the relation $E_{\rm IMT}(T_{\rm b})$ displays an exponential dependence with the bath
temperature close to $T_{\rm N}$, and therefore a rapid decrease of $E_{\rm IMT}(T_{\rm b})$ near $T_{\rm b}\approx T_{\rm N}$ is expected.

\subsubsection{Finite-Temperature MIT}
The temperature dependence of the MIT is easier to analyze. Since the MIT
concerns the stability of the metallic phase, the effective temperature
relation derived in Eq.~(\ref{tefftb}) holds at the MIT. 
This yields
\begin{align} \label{mitt}
E_{\rm MIT}(T_{\rm b}) = \sqrt{\frac23} \frac{\pi \Gamma}{v_0} \sqrt{T_{\rm N}^2 - T_{\rm b}^2},
\end{align}
and at high temperatures close to $T_{\rm N}$, it yields the scaling relation
\begin{align}
E_{\rm MIT}(T_{\rm b}) \simeq  E_{\rm MIT}(0) \sqrt{\frac{T_{\rm N} -T_{\rm b}}{T_{\rm N}} }.
\end{align}
Our theory thus successfully reproduces the square-root behavior $E_{\rm MIT}(T_{\rm b})$ near $T_{\rm N}$ which had been observed experimentally~\cite{tlwu}.
Numerically, the same procedure as described in Appendix~\ref{sec:mit} can be used, computing the off-diagonal GF in Eq.~(\ref{gapeq})  as
\begin{equation}
-\int_{-\infty}^\infty \frac{2\Gamma\Delta\omega}{v_0^4(k_1^2+k_2^2)}
\rme^{-2k_2|x|}f_0(\omega+Ex)\rmd x,
\end{equation}
with $k_{1,2}$ given in Eq.~(\ref{k1k2}) and the Fermi-Dirac function
$f_0(\omega)$ at temperature $T_{\rm b}$. The oscillatory parts of the
off-diagonal GF are ignored in this calculation.
The numerical results for $E_{\rm MIT}(T_{\rm
b})$ are shown with red circles, validating furthermore  the above square-root scaling
relation.

The mean-field theory seems to predict a very slow decrease of $E_{\rm IMT}(T_{\rm b} \ll T_{\rm N} )$, as depicted in
FIG.~\ref{fig:evst}. Consequently, it predicts a wide bi-stability region where both metallic and insulating phases can coexist. 
One has to keep in mind that a mean-field approach typically exaggerates the domain of stability of ordered states,
and only a more sophisticated diagrammatic theory could resolve this issue.
 We emphasize that the bath-temperature
dependence at the RS is significant even when the underlying mechanism
is electronic, and a large reduction of the switching field over an order
of $\Gamma/\Delta_0$ should be carefully taken into interpretation when the
energy scale of switching field is examined.

\section{Towards an Effective Field Theory of Resistive Switching}

\begin{figure}
\begin{center}
\rotatebox{0}{\resizebox{1.5in}{!}{\includegraphics{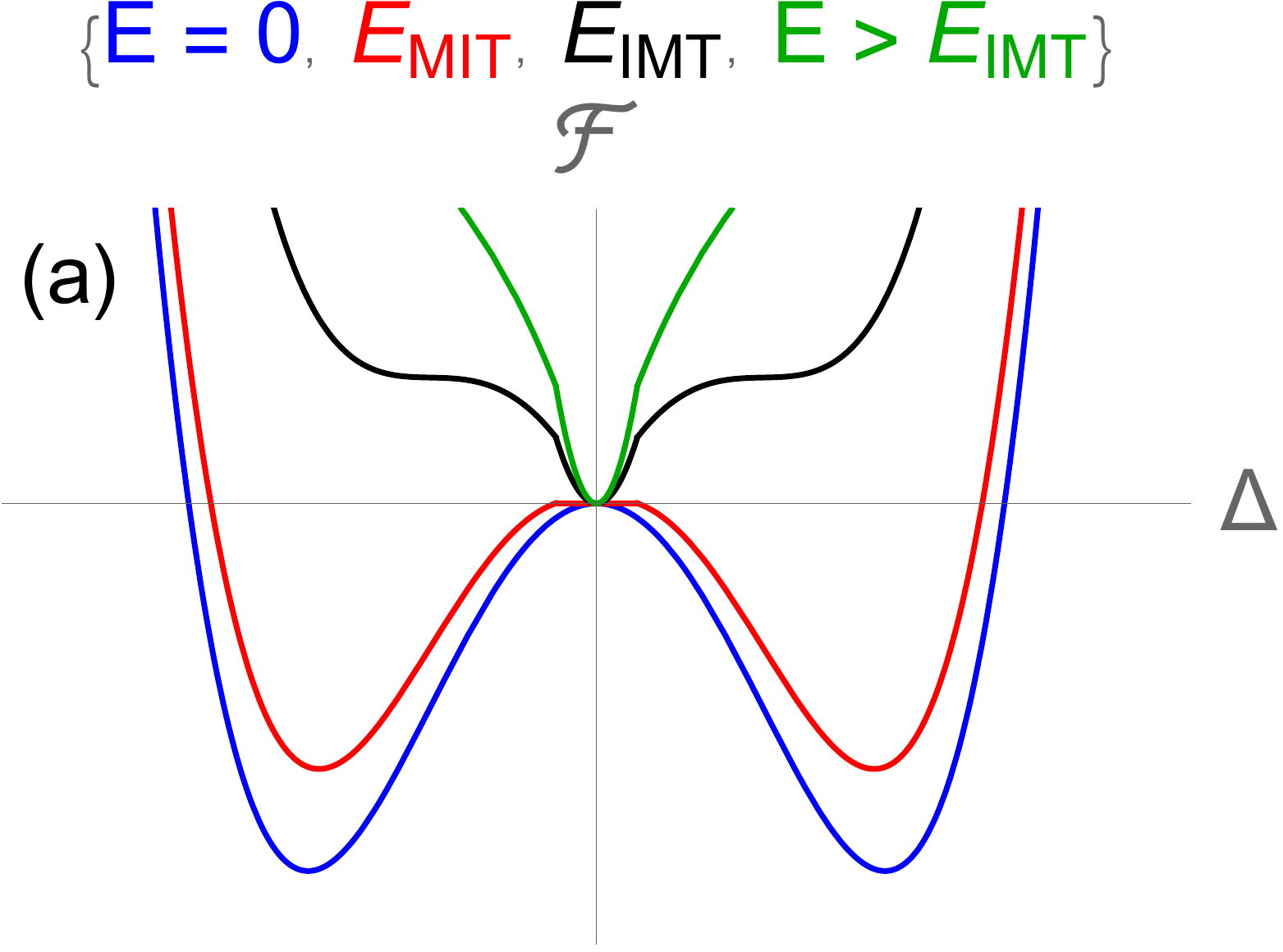}}}
\hspace{1em}
\rotatebox{0}{\resizebox{1.5in}{!}{\includegraphics{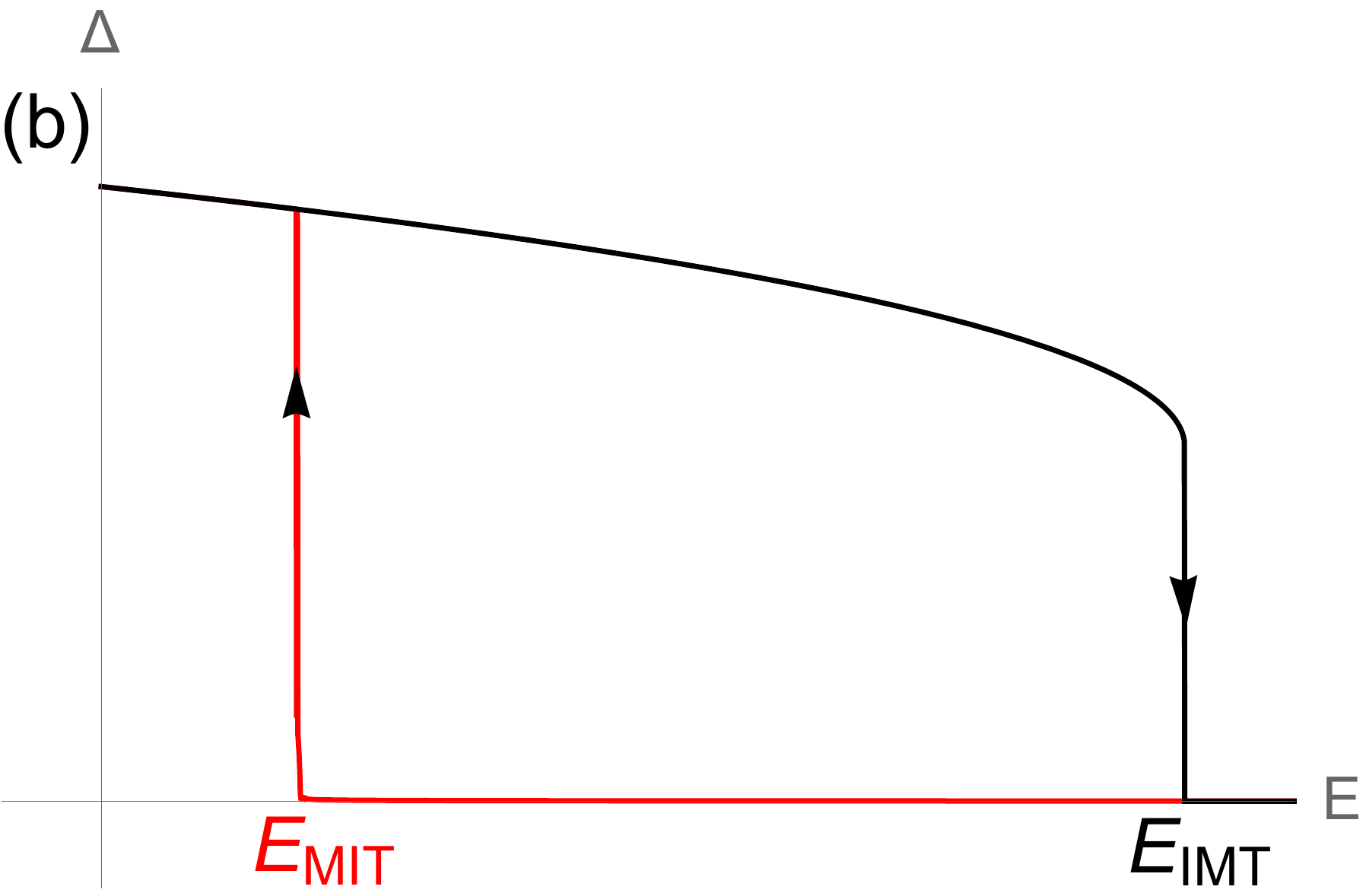}}}
\caption{(a) Shape of the effective free energy $\mathcal{F}(\Delta)$ proposed in Eq.~(\ref{eq:freeenergy}) when varying the electric field $E$. (b) Corresponding spontaneous order parameter $\Delta$, when varying $E$. The hysteresis, and the bi-stability region between the MIT and the IMT, emerge naturally from the two regimes of effective temperature given in Eq.~(\ref{eq:teffboth}).
}
\label{fig:free}
\end{center}
\end{figure}

In this Section, we leverage the teachings of the previous mean-field
analysis to propose a low-energy effective theory description of the
local order parameter $\Delta$ at both the MIT and the IMT.  This could
provide a practical path to developing an effective \emph{field} theory
capturing the spatial fluctuations of the order parameter, which are
critical to the understanding of realistic resistive switching
phenomena.  The problem being far from equilibrium, such an effective
theory should not only determine the gap $\Delta$ (a spectral quantity
obtained from $G^r$), but also the nonequilibrium excitations (such as
the quantity $T_{\rm eff}$, obtained from the ratio of $G^<$ and $G^r$).
In principle, only a fully nonequilibrium approach such as the quantum
Schwinger-Keldsyh formalism or the classical Martin-Siggia-Rose
formalism~\cite{msr,schmittmann} can tackle both order parameters,
$\Delta$ and $T_{\rm eff}$, on an equal footing.  However, we aim at a
simpler description by constructing an effective Ginzburg-Landau free
energy for $\Delta$ alone, $\mathcal{F}(\Delta)$, under a finite
electric field.  Instead of being a dynamical quantity, the effective
temperature will be fixed by using an educated ansatz, $T_{\rm
eff}(\Delta)$, based on the results of the previous Sections.  Although
certainly less rigorous than a Schwinger-Keldsyh or Martin-Siggia-Rose
approach, this static approach does not require solving time dynamics,
giving a huge computational advantage when extending the theory to large
heterogeneous systems including phase segregation~\cite{nanolett}.

The functional form of ${\cal F}(\Delta)$ is dictated by the $\mathbb{Z}_2$ symmetry of the order parameter, and its minima and their stability are imposed by the gap equation,
\begin{align}
&\mbox{zeros}\left\{  \frac{\rmd{\cal F}(\Delta)}{\rmd\Delta} \right\} = \nonumber \\
& \
\mbox{zeros} \left\{ \Delta-
U\int_{-D}^D \!\frac{\rmd\omega}{2\pi\rmi}  \left[G^<_{-+}(\omega)+G^<_{+-}(\omega)\right] \right\},  \\
&\mbox{sign}\left\{  \frac{\rmd^2{\cal F}(\Delta)}{\rmd\Delta^2} \Big{|}_{\mbox{zeros}} \right\} =  \\
& \
\mbox{sign} \left\{ 1 -
U\int_{-D}^D  \! \frac{\rmd\omega}{2\pi\rmi}   \frac{\rmd }{\rmd \Delta} \left[G^<_{-+}(\omega)+G^<_{+-}(\omega)\right] \Big{|}_{\mbox{zeros}}\right\}. \nonumber
\end{align}
Another constraint on ${\cal F}(\Delta)$ comes from the equilibrium limit ($E=0$) for which the Ginzburg-Landau free energy is an Ising $\phi^4$-theory reading
\begin{equation}
\mathcal{F}_{\rm eq}(\Delta) = (T - T_{\rm N}) \, \Delta^2 + \lambda \, \Delta^4 + \ldots
\end{equation}
where $T$ is the temperature of the system, and $T_{\rm N}$ is the N\'eel temperature at which the equilibrium transition occurs.
The interaction parameter $\lambda > 0$ can be set by requiring that $\Delta = \Delta_0$ at zero temperature, yielding $\lambda \approx T_{\rm N}/2 \Delta_0^2$.
All these constraints lead us to propose the following effective Ginzburg-Landau free energy
\begin{align} \label{eq:freeenergy}
\mathcal{F}(\Delta) = \left( T_{\rm eff}(\Delta) - T_{\rm N} \right) \Delta^2 + \lambda \, \Delta^4 + \ldots
\end{align}
with the state-dependent effective temperature given, at zero bath temperature, by the expressions in Eqs.~(\ref{eq:teff0}) and (\ref{eq:teff1}), 
\begin{align} \label{eq:teffboth}
T_{\rm eff}(\Delta)  \sim \left\{ 
\begin{array}{l}
 v_0 E /\Gamma \mbox{ \rmi n the small } \Delta \mbox{ regime,} \\
 v_0 E / |\Delta|  \mbox{ \rmi n the large } \Delta \mbox{ regime.}
\end{array}
\right.
\end{align}
Importantly, the electric field now enters the
problem solely through the renormalization of the temperature to a
gap-dependent effective temperature $T_{\rm eff}(\Delta)$. This
constitutive relation is the only remainder of the nonequilibrium nature
of the problem.  The distance to the N\'eel temperature $T_{\rm N}$ in
the $\Delta^2$ term controls the stability of the metallic phase at
$\Delta = 0$. In the spirit of an effective field-theory description, 
the coefficients of the higher order terms are expected
to play an irrelevant role near the transitions, merely renormalizing
the transition temperatures.

Interestingly enough, in the large $\Delta$ regime, the electric field
couples to the order parameter linearly via the non-analytic term $E
|\Delta|$, which transforms the continuous equilibrium phase transition
into a discontinuous resistive switching. In this Ginzburg-Landau
language, the MIT and IMT correspond to the destabilization of a
metastable solution, \textit{i.e.} to the disappearing of a local
minimum  of $\mathcal{F}(\Delta)$ to the profit of a global minimum.
For example, at the IMT the insulating solution at $\Delta \approx
\Delta_0/\sqrt{3}$ is destabilized when the effective temperature
reaches 
\begin{equation}
T_{\rm IMT} \approx \frac{4}{3} T_{\rm N},
\label{timt_fe}
\end{equation}
which is naturally
consistent with our previous findings, see Eq.~(\ref{timt}), up to small
differences in the numerical factors due to the truncation of the free
energy to lowest orders. The conclusion that $T_{\rm IMT}$ is controlled by $T_{\rm N}$ is valid
regardless of the precise $E$-field dependence in
Eq.~(\ref{eq:teffboth}) as long as $T_{\rm eff}(\Delta)\propto 1/\Delta$ at large $\Delta$.

FIG.~\ref{fig:free} sketches the evolution of the shape of
$\mathcal{F}(\Delta)$, when increasing $E$ starting from a stable
insulating state ($\Delta > 0$), rapidly developing a second stable
minimum at $\Delta = 0$ which becomes the only stable minimum at $E_{\rm
IMT}$, when the insulating state becomes unstable. When decreasing the
electric field from this metallic state, the stability of the latter is
lost  at a much lower electric field $E_{\rm MIT} \sim (\Gamma/\Delta_0)
E_{\rm IMT}$.

\section{Conclusions}

We have worked out an analytic window into the inner workings of  RS in
correlated insulators close to an equilibrium phase transition by means
of a mean-field (MF) treatment of a minimal model of a
driven-dissipative anti-ferromagnet.  This allowed to unambiguously
resolve the age-old debate on whether the RS is mainly electronically
driven or thermally driven: both scenarios were reconciled in a unified
picture where the nonequilibrium electronic excitations were
characterized by a state-dependent effective temperature $T_{\rm eff}$.
While the underlying physical mechanism responsible for the latter is
different in the insulating state (mostly electronic Landau-Zener
events) from that of the metallic state (mostly thermal heating caused
by the dissipative mechanisms), both the IMT and the IMT were shown to
occur whenever $T_{\rm eff}$ reaches $T_{\rm N}$, the equilibrium N\'eel
transition temperature.  Concomitantly, our analytics also provided an
elegant resolution to the puzzle posed by the disconcertingly small
threshold fields when compared to the typical spectral energy scales:
the electric field does not affect substantially the spectrum of the
materials, but enters the problem through the effective temperature
$T_{\rm eff}$. While the latter is comparable to the bandgap at the RS,
the electric field can be orders of magnitude smaller.

While the analytic MF approach makes the theory transparent, the range
of validity of the MF approximation in nonequilibrium situations is
largely untested. Although the agreement with many salient experimental
features is very encouraging~\cite{nanolett}, our theoretical approach
can only be taken as an initial reference  point in the construction of
a more comprehensive theory of RS. Given the existence of a bi-stability
region between the IMT and the MIT, the possibility for the system to
develop spatial inhomogeneities is a crucial element of the
resistive-switching transition~\citep{nanolett}.  Experimental and
numerical studies revealed that the electron conduction is often carried
through metallic filaments and the details of the $I$-$V$ characteristic
strongly depends on the filament dynamics. Therefore, the next step of
this program should be to question the influence of spatial fluctuations
on the critical points by upgrading the above Ginzburg-Landau free
energy to a full-fledged functional of the order parameter field
$\Delta(\boldsymbol{x})$, and perform a renormalization-group treatment.
Another important step should be to investigate the role of the
fluctuations around the mean-field solution, classically and quantum
mechanically.  Finally, exploring diverse RS phenomena to guide the
design of possible devices will require improving the numerical
methodologies in order to perform realistic calculations of
material-specific models.

\begin{acknowledgements}
JEH acknowledges the computational support from the CCR at University at
Buffalo. We thank Martin Eckstein, Sambandamurthy Ganapathy, P\'ia Homma
Jara, Hyun-Tak Kim, Alfred Leitenstorfer, Steve Leone, Mariela Menghini,
Danny Shahar, and Philipp Werner for helpful discussions.
\end{acknowledgements}

\vfill

\pagebreak

\appendix

\section{Numerical Calculation of Wavefunction}
\label{app:wf}

The parabolic cylinder function~\cite{whittaker,gradshteyn} in
Eq.~(\ref{pm}) can be expressed as
\begin{eqnarray}
D_p(z) & = &
2^{p/2}\rme^{-z^2/4}\left[\frac{\sqrt{\pi}}{\Gamma\left(\frac{1-p}{2}\right)}\Phi\left(
-\frac{p}{2},\frac12;\frac{z^2}{2}\right)\right. \nonumber \\
& &
-\left.\frac{\sqrt{2\pi}z}{\Gamma\left(\frac{-p}{2}\right)}\Phi\left(
\frac{1-p}{2},\frac32;\frac{z^2}{2}\right)
\right],
\label{dpdef}
\end{eqnarray}
with the confluent hypergeometric function $\Phi(a,b;z)$.
The equality $\Phi(a,b;0)=1$ is useful.
Directly computing the parabolic cylinder function numerically from the
hypergeometric function, however, turns out very unreliable, especially with a
complex index $i\alpha/2$. Instead, we obtain the solution to the Hamiltonian
by integrating the differential equation, Eq.~(\ref{diff1}). Since we
expect rapid oscillations due to the electrostatic potential,
we absorb the fast oscillation as
\begin{equation}
\phi^R_+(x)=a(x)\rme^{\rmi(\omega x+\frac12 Ex^2)}\mbox{ and }
\phi^R_-(x)=b(x)\rme^{-\rmi(\omega x+\frac12 Ex^2)}.
\end{equation}
with the differential equations
\begin{eqnarray}
-\rmi a'(x) & = & \Delta b(x)\rme^{-2\rmi(\omega x+\frac12 Ex^2)} \\
\rmi b'(x) & = & \Delta a(x)\rme^{2\rmi (\omega x+\frac12 Ex^2)}.
\end{eqnarray}
Since $x$ and $\omega$ always appear as $x+\omega/E$, one only needs to
compute for $\omega=0$ and later translate $x\to x+\omega/E$ at any
non-zero $\omega$.
Setting the boundary condition is crucial to produce the physical
solution and avoid any divergent results. 
The best method is to set the wavefunction values at $x=0$ by
Eqs.~(\ref{phiR+}), (\ref{phiR-}) and
$D_p(0)=2^{p/2}\sqrt{\pi}/\Gamma(\frac{1-p}{2})$ from Eq.~(\ref{dpdef}), and
integrate the equations outwards to $\pm\infty$.

The local spectral weight in the limit $\Gamma\to 0$ can be
evaluated from Eq.~(\ref{retardg}) as
$\rho_{++}(0,0;\omega)+\rho_{--}(0,0;\omega)$. At $\omega=0$, only
the imaginary part is non-zero for $G^r_{\rm loc}(0)$ and
$\mbox{Im }\,G^r_{\rm loc}(0)=-(4
v_0)^{-1}[|\phi^R_+(0)|^2+|\phi^L_+(0)|^2
+|\phi^R_-(0)|^2+|\phi^L_-(0)|^2]=-(2
v_0)^{-1}[|\phi^R_+(0)|^2+|\phi^L_+(0)|^2]$.
Using the definition of the
parabolic cylinder function~\cite{whittaker,gradshteyn}, $-\mbox{Im }\,G^r_{\rm
loc}(0)=(2v_0)^{-1}\rme^{-3\pi\alpha/4}[|D_{-\rmi\alpha/2}(0)|^2
+\frac{\alpha}{2}|D_{-\rmi\alpha/2-1}(0)|^2]
=(2v_0)^{-1}\rme^{-\pi\alpha/2}$. As shown in the
inset of FIG.~\ref{fig:spec}(a), the spectral weight at $\omega=0$ decays
exponentially with $\alpha$ until the damping-induced in-gap weight
becomes more dominant.

\section{Integrals for Retarded GF in the Landau-Zener Regime}
\label{app:intLZ}

\begin{figure}
\begin{center}
\rotatebox{0}{\resizebox{3.0in}{!}{\includegraphics{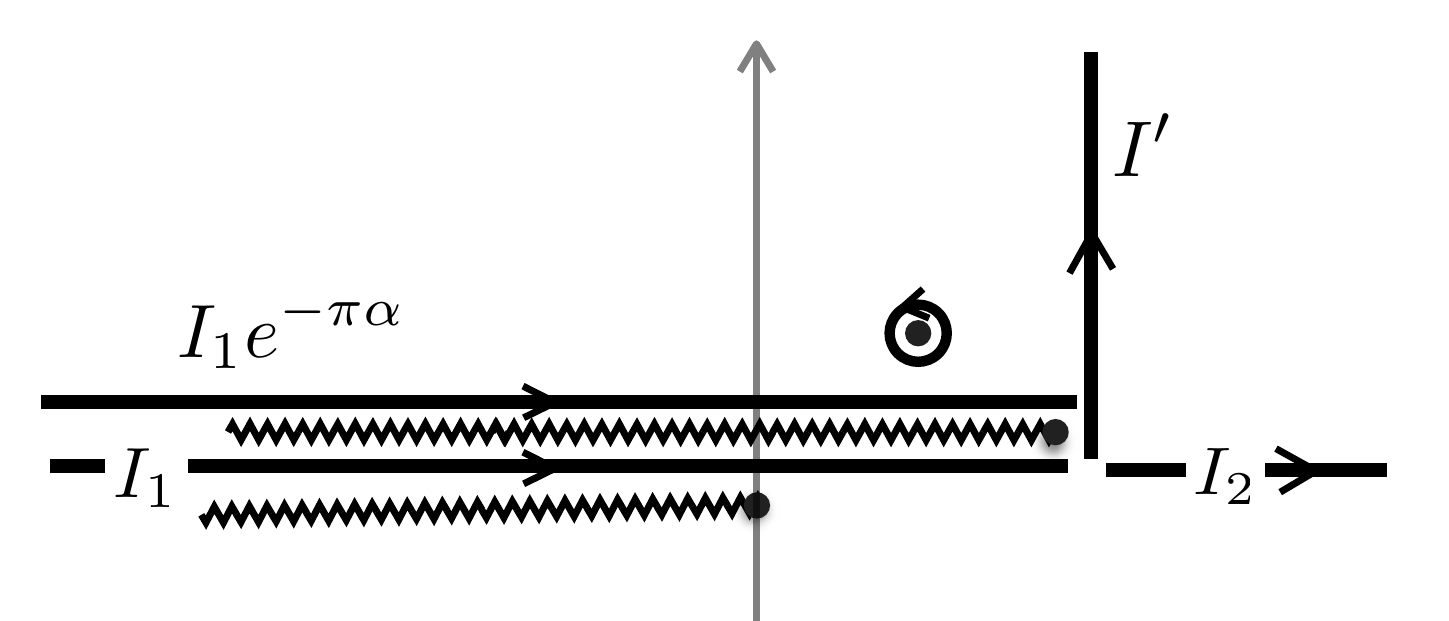}}}
\caption{Contour for the integral Eq.~(\ref{contourint}).
}
\label{fig:cont1}
\end{center}
\end{figure}

From Eq.~(\ref{lesserg}), $x<0$ has finite contribution for
$\omega>0$, and with this the first exponential factor decays as we
enclose the $\omega'$-contour in the upper-half-plane. To obtain the
integral~(\ref{finalgr}) we organize the integral contour as shown in
FIG.~\ref{fig:cont1}. The desired integral is $I_1+I_2$. The integral
$I_2$ can be rotated to $I'$ which converges much faster than $I_2$
due to the exponentially decaying factor in
$\rme^{-\rmi E/(2v_0)(x^2+2x\omega'/E)}$ for $x<0$ and $\omega'=\rmi y$ ($y>0$).
The integral above the contour $C_2$ with the contribution
$I_1\rme^{\pi\alpha}$ combines with $I'$ to give the residue integral 
at $\omega'=\omega+\rmi\Gamma$,
\begin{equation}
I_1\rme^{-\pi\alpha}+I'=-(\rmi /v_0) \rho_{++}(0,x;\omega+\rmi\Gamma).
\end{equation}
Therefore $G^r_{++}(0,x;\omega)$ can be expressed as
\begin{equation}
\rme^{\pi\alpha}[-(\rmi /v_0)\rho_{++}(0,x;\omega+\rmi\Gamma)-I']+I'.
\end{equation}
The term proportional to the residue becomes
\begin{equation}
-\frac{\rmi}{v_0}\rme^{-\frac{\rmi E}{2v_0}(x^2+\frac{2x(\omega+\rmi\Gamma)}{E})}
\left|\frac{\omega+\rmi\Gamma}{\omega-E|x|+}
\right|^{-\rmi\alpha/2}\rme^{-\alpha\varphi/2}.
\end{equation}
The remaining term $(\rme^{\pi\alpha}-1)I'$ can be easily evaluated due to
the contour rotation in $I'$, 
\begin{equation}
\int_0^{\infty}\frac{\rme^{-|x|y}\left|\frac{E|x|+\rmi y}{\rmi y}\right|^{-\rmi\alpha/2}
\rme^{\alpha/2(\tan^{-1}(y/E|x|)-\pi/2)}}{
\omega-E|x|+\rmi y+\rmi\Gamma}\frac{\rmi \rmd y}{2\pi},
\end{equation}
whose integral range is set by $|x|^{-1}$ and the integral is then well
approximated by
\begin{equation}
\frac{-\rmi\rme^{-\pi\alpha/4}}{2\pi(E|x|-\omega)|x|}
(Ex^2)^{-\rmi\alpha/2}\Gamma\left(1+\rmi\frac{\alpha}{2}\right).
\end{equation}
In the small damping and $\alpha$ limit, the residue contribution dominates
$(\rme^{\pi\alpha}-1)I'$
and we arrive at Eq.~(\ref{finalgr}).

\section{Small-Field Approximation}
\label{sec:zerofield}

In the limit of large $\alpha$ with $E\ll\Delta$, approximating the
retarded GF by that of zero-field limit may be a reasonable
approximation. The justification of this idea is discussed further in the next
section. The calculation of the lesser GF, however, is done with the
full
nonequilibrium Dyson's equation (\ref{lesserg}).
The zero-field retarded GF can be
written down in the A/B sublattice basis as
\begin{equation}
{\bf G}^r(\omega,p)=\left(
\begin{array}{cc}
\omega+\Delta+\rmi\Gamma & -v_0p \\
-v_0p & \omega-\Delta+\rmi\Gamma
\end{array}
\right)^{-1},
\end{equation}
with the momentum $p$.
Therefore, the retarded GF on the $A$-sublattice is given as
\begin{eqnarray}
G^r_{AA}(x,\omega) & = & \int_{-\infty}^\infty\frac{(\omega-\Delta+\rmi\Gamma)\rme^{\rmi px}}{
(\omega+\rmi\Gamma)^2-\Delta^2-v_0^2p^2}\frac{dp}{2\pi}
\nonumber \\
& = &
 -\frac{\rmi}{2v_0^2}\frac{\omega-\Delta+\rmi\Gamma}{k_1+\rmi k_2}\rme^{ \rmi  k_1 x
-k_2|x|},\nonumber \\
G^r_{AB}(x,\omega) & = & 
\frac{\rmi}{2v_0}\rme^{\rmi k_1 x
-k_2|x|},
\end{eqnarray}
with
\begin{eqnarray}
v_0k_2 & = & \left[
\left(u^2+\omega^2\Gamma^2\right)^{1/2} -u
\right]^{1/2}\nonumber \\
v_0^2k_1k_2 & = & \omega\Gamma\mbox{ and }
u=(\omega^2-\Delta^2-\Gamma^2)/2.
\label{k1k2}
\end{eqnarray}
Then, the local lesser GF at $x=0$, Eq.~(\ref{lesserg}), is rewritten a
\begin{equation}
G^<_{AA}(\omega)=2\rmi\Gamma\int_{-\infty}^{-\omega/E}\left[
|G^r_{AA}(x,\omega)|^2
+|G^r_{AB}(x,\omega)|^2\right]\rmd x.
\label{app:gless}
\end{equation}
After straightforward calculations, one obtains
\begin{eqnarray}
G^<_{AA}(\omega) & = & \rmi
\frac{(\omega-\Delta)k_1+\Gamma k_2}{v_0^2(k_1^2+k_2^2)}
f(\omega)\nonumber \\
&= & 2\rmi\pi\left(-\frac{1}{\pi}\mbox{Im }G^r_{AA}(0,\omega)
\right)f(\omega)
\end{eqnarray}
with the distribution function
\begin{eqnarray}
f(\omega) & = & 
\frac12\exp\left(-\frac{2k_2(\omega)\omega}{E}
\right)\Theta(\omega) \nonumber \\
& + & 
\left[1-\frac12\exp\left(\frac{2k_2(\omega)\omega}{E}
\right)\right]\Theta(-\omega).
\end{eqnarray}
The distribution function assumes the same form as the free 1-$d$ model,
Eq.~(\ref{feff0}) with the inverse penetration depth $k_2(\omega)$
replacing $\Gamma$. In the $\Delta\to 0$ limit, $k_2$ becomes $\Gamma$.
With a finite field with $|\omega|<\Delta$, the gap acts like a
potential barrier and the wavefunction decays under the gap with the rate proportional
to $\Delta$, leading to $k_2(\omega)\simeq \Delta/v_0$. Therefore the
distribution function and the retarded GF are expected to behave as
\begin{eqnarray}
f(\omega) & \simeq & \frac12 \rme^{-2\Delta\omega/v_0E}
\nonumber \\
-\frac{1}{\pi}\mbox{Im }G^r_{AA}(0) & = & \frac{1}{2\pi v_0^2}\frac{\Gamma}{k_2}
= \frac{\Gamma}{2\pi v_0\Delta}
\end{eqnarray}
for $0<\omega\lesssim \Delta$
in the large $\alpha$ limit, as verified in FIG.~\ref{fig:spec}.

For the gap equation, the order parameter is calculated
self-consistently as
\begin{eqnarray}
\Delta & = & \frac{U}{2}(n_A-n_B) 
\nonumber \\
& = & \frac{U}{2}\int [G^<_{AA}(\omega;\Delta)
-G^<_{AA}(\omega;-\Delta)]\frac{\rmd\omega }{2\pi\rmi}
\nonumber \\
& = & -\Delta\frac{U}{2\pi v_0}\int\frac{k_1 f(\omega)}{v_0(k_1^2+k_2^2)}
\rmd\omega.
\label{gapapp}
\end{eqnarray}

\section{Off-Diagonal Green's Functions}
\label{sec:goff}

The spectral function of the off-diagonal retarded GFs
$G^r_{+-}(0,0;\omega)+G^r_{-+}(0,0;\omega)$, responsible for the gap
equation, is given from
Eq.~(\ref{spectral}) in the small $\Gamma$ limit as
\begin{equation}
2{\rm
Re}\left\{\phi_+^R(0,\omega)[\phi_-^R(0,\omega)]^*
-\phi_+^R(0,-\omega)[\phi_-^R(0,-\omega)]^*\right\}
\label{goffspec}
\end{equation}
after using the symmetry relations
Eq.~(\ref{symmetry}). This spectral function is purely real and
odd in $\omega$. In the small-$\omega$ limit, the exact definition of the
wavefunction (\ref{phiR+}) and (\ref{phiR-}) with Eq.~(\ref{dpdef}) can
be used to expand the spectral function in the lowest order of $\omega$
as
\begin{equation}
-\frac{2\sqrt{2}}{3}\frac{\Delta
\rme^{-\pi\alpha}}{(v_0E)^2}\omega^3\mbox{ for }\omega\simeq 0.
\end{equation}
with the spectral weight suppressed by the LZ factor inside the gap.

In the large-$\omega$ limit, the asymptotic expansion (\ref{phiR})
can be used to
evaluate the GF. The wavefunction $\phi_\pm^R(0,\omega)$ consists of
three contributions away from $\omega=0$: incoming, transmitted and
reflected waves. For instance, for $\omega>0$ $\phi_+^R(0,\omega)$ has
the transmitted wave. Due to the oscillation $\rme^{\pm \rmi y^2}$ in
Eq.~(\ref{phiR}) induced by the external field, the product between the wavefunction components may have
cancelled or strong phase oscillations. For example, a product of
incoming waves in $\phi_+^R(0,-\omega)$ and $\phi_-^R(0,-\omega)$ of
Eq.~(\ref{goffspec}) has
the most dominant contribution that has cancelled phases.
Cross-component products have uncancelled phases as $\rme^{\pm
\rmi\omega^2/v_0E}$. The oscillations become more rapid for smaller electric
field. Such strong oscillation present in both frequency $\omega$ and position
$x$ makes the numerical calculations quite challenging at small fields.

\begin{figure}
\begin{center}
\rotatebox{0}{\resizebox{3.0in}{!}{\includegraphics{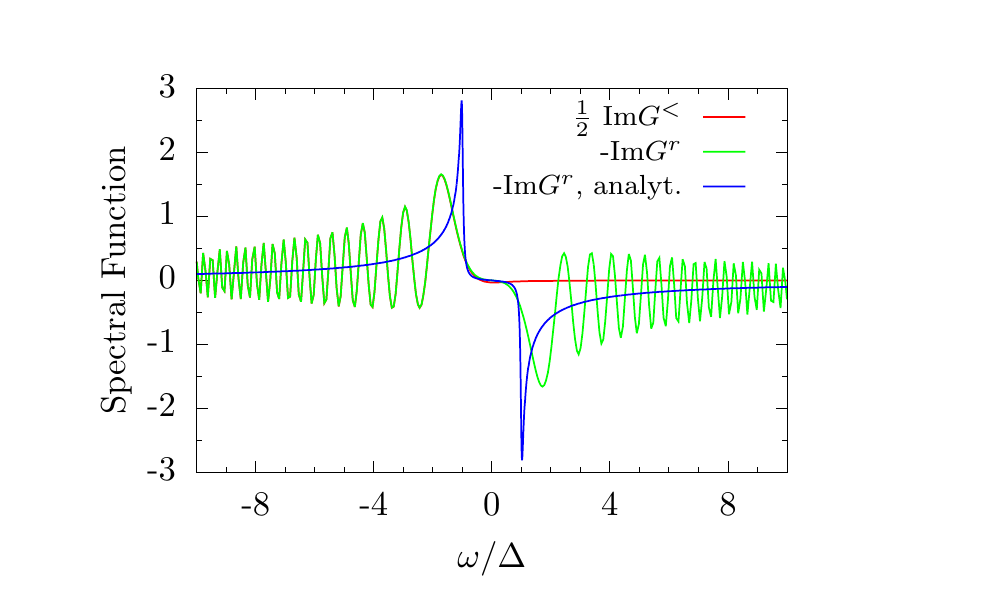}}}
\caption{
Off-diagonal GFs $G^{r,<}_{+-}(\omega)+G^{r,<}_{-+}(\omega)$ at
$\Delta=\Delta_{\rm IMT}$, $E=E_{\rm IMT}$
and $\Gamma=0.01$. Fully numerical calculations for the lesser (red) and
retarded (green) GFs show strong oscillation in frequency. The center of
oscillation is well described by the analytic evaluation (blue) of the
non-oscillatory part in the zero-field GF.
}
\label{fig:g12spec}
\end{center}
\end{figure}

Analytic calculation for the non-oscillatory contribution at
$|\omega|\gtrsim \Delta$ gives the approximate expression~\cite{gradshteyn}
\begin{equation}
-\frac{\Delta}{\omega}\left(1+\frac{\Delta^2}{2\omega^2}\right)
 \mbox{ for }|\omega|\gtrsim \Delta\mbox{ and }
\frac{\Delta^2}{v_0E}\gtrsim 1,
\label{nonoscill}
\end{equation}
which is, except for the prefactor $\Delta$, the same as the large-$\omega$ expansion of
$(\omega^2-\Delta^2)^{-1/2}$ in the gap equation~(\ref{gapimt}).
It is remarkable that the non-oscillatory part of the integral is
independent of the electric field. FIG.~\ref{fig:g12spec} shows the
the zero-field retarded
GF (blue) threads the center of oscillation of the numerically accurate retarded
GF (green). Although the integration of the oscillatory part is
non-zero, especially with the important contribution close to
$|\omega|\sim\Delta$, the approximation by the zero-field retarded GF is
reasonable.

\section{Switching Field at the MIT}
\label{sec:mit}

The on-set of the MIT is determined by the stability of the $\Delta=0$
solution in Eq.~(\ref{gapeq}), the condition that the slope of the {\sc rhs}
remains below 1. Therefore the condition for the MIT is
\begin{equation}
\frac{1}{U}=\lim_{\Delta\to 0}\frac{1}{\Delta}\int
\left[G^<_{-+}(\omega)+G^<_{+-}(\omega)\right]\frac{\rmd\omega}{2\pi\rmi}.
\end{equation}
We can evaluate this exactly by using the first-order expansion of the
GF out of the non-interacting GF considered in Section~\ref{sec:metal}. The
retarded GF satisfies
\begin{equation}
(\omega+\rmi\Gamma-\hat{H}_0){\bf G}^r(x,x')
=\delta(x-x'){\bf I},
\end{equation}
with the GF matrix $({\bf G}^r)_{ab}=G^r_{ab}$. $\hat{H}_0$ acts on the
$x$. Taking the first-order expansion gives 
\begin{equation}
\Delta G^{r,0}_{++}(x,x')+(\omega+\rmi\Gamma-\rmi v_0\partial +Ex)G^r_{-+}(x,x')=0.
\end{equation}
Here, we suppressed $\omega$ in the expression for brevity.
The unperturbed GF $G^{r,0}_{++}(x,x')$ is given in Eq.~(\ref{g+0}) with
$\lambda=+$. Defining $g(x,x')=\rme^{\rmi\varphi(x,x')-(\Gamma/v_0)|x-x'|}G^r_{-+}(x,x')$, one
solves the differential equation to obtain for $x>x'$ as
\begin{equation}
g(x,x')=g_0(x')-\frac{\Delta}{v_0^2}\int_{x'}^x
\rme^{2\rmi\varphi(y,x')-2(\Gamma/v_0)(y-x')} \rmd y
\end{equation}
with an arbitrary function $g_0(x')$. Since $g(x,x')\to 0$ as
$|x-x'|\to\infty$, one sets the boundary condition as
\begin{equation}
g(x,x')=\frac{\Delta}{v_0^2}\int^{\infty}_x \rme^{-2(\Gamma/v_0)(y-x')+2\rmi\varphi(y,x')}\rmd y.
\end{equation}
This gives us for $x<0$
\begin{equation}
G^r_{-+}(0,x)=\frac{\Delta}{v_0^2} \rme^{(\Gamma/v_0) x+\rmi \varphi(0,x)}\int_0^\infty
\rme^{-2(\Gamma/v_0) y+2\rmi\varphi(y,0)} \rmd y,
\end{equation}
and 
\begin{equation}
G^r_{-+}(0,x)G^{r,0}_{++}(0,x)^*
=\rmi\frac{\Delta}{v_0^3}\Theta(-x)\rme^{2(\Gamma/v_0) x}I(\omega,\Gamma,E),
\end{equation}
with the integral denoted as $I(\omega,\Gamma,E)$. Performing an
integral over $x$ in Eq.~(\ref{lesserg}), we get
\begin{equation}
v_0^2G^<_{-+}(\omega)/\Delta=-I(\omega,\Gamma,E)\times\left\{
\begin{array}{ll}
\rme^{-2\Gamma\omega/v_0E} & \omega>0 \\
1 & \omega<0
\end{array}\right..\nonumber
\end{equation}
Similarly one obtains
\begin{equation}
v_0^2G^<_{+-}(\omega)/\Delta=-I(\omega,\Gamma,-E)\times\left\{
\begin{array}{ll}
0 & \omega>0 \\
1-\rme^{-2\Gamma\omega/v_0E} & \omega<0
\end{array}\right..\nonumber
\end{equation}
The integral $I(\omega,\Gamma,E)$ can be transformed to the 
parabolic cylinder function $D_{-1}(z)$ with
$z=\sqrt{2/v_0E}\rme^{-\rmi\pi/4}(\omega+\rmi\Gamma)$
by rotating the integration contour, and then approximated by the
asymptotic expansion as
\begin{equation}
\frac{\rmi v_0}{2(\omega+\rmi\Gamma)}
+\Theta(-\omega)\frac{\rme^{\rmi\pi/4}}{\sqrt{E/\pi v_0}}\exp\left[
\frac{-\rmi(\omega+\rmi\Gamma)^2}{v_0E}\right].
\end{equation}
$I(\omega,\Gamma,-E)$ can be obtained by replacing $E\to \rme^{\rmi\pi}E$ and
$\Theta(-\omega)\to\Theta(\omega)$. The second term is highly
oscillatory and we ignore its integral in the analytic estimate. It also
shows that the oscillation goes like $\rme^{-\rmi\omega^2/v_0E}$ for large
$\omega$ and why the
numerical calculation becomes problematic in the small $E$ limit.
Then the first term
is nothing but the zero-field retarded GF.
Combining the results, we arrive at the MIT condition
\begin{equation}
\frac{2\pi v_0}{U}=-\int^D_{-D}\frac{\omega f^0_{\rm eff}(\omega)}{\omega^2+\Gamma^2}
\rmd\omega
\end{equation}
with the non-interacting distribution function Eq.~(\ref{feff0}).
Performing this integral in a similar manner as considered in the main
text, we obtain the integral
\begin{equation}
\int_0^D\frac{\omega \rmd\omega}{\omega^2+\Gamma^2}
-\int_0^\infty\frac{\omega \rme^{-2\Gamma\omega/v_0E}\rmd\omega}{\omega^2+\Gamma^2}
\simeq
\ln\left(\frac{2\rme^\gamma\Gamma D}{v_0 E}\right)
\end{equation}
in the limit $\Gamma \ll T_{\rm eff}\sim v_0 E / \Gamma \ll D $.
This analytic expression agrees very well with the exact value by
numerically evaluating $I(\omega,\Gamma,E)$ within 5\%
for the parameters considered.
We then have the MIT condition at the switching field $E_{\rm MIT}$ as
\begin{equation}
E_{\rm MIT}\simeq \rme^\gamma\frac{\Gamma}{v_0}\Delta_0\approx
1.78\,\frac{\Gamma}{v_0}\Delta_0,
\end{equation}
which gives, with the parameters used in this work, the analytic estimate 
$E_{\rm MIT}=0.0053$ at about 20\% overestimate from the numerical
value.


%

\end{document}